\documentclass[longauth]{aa}

\usepackage{txfonts}
\usepackage{graphicx}
\usepackage{natbib}
\bibpunct{(}{)}{;}{a}{}{,}

\begin{document}

\title{The VIMOS VLT Deep Survey}
\subtitle{The contribution of minor mergers to the growth of $L_{B} \gtrsim L_{B}^{*}$ galaxies 
since $z \sim 1$ from spectroscopically identified pairs}

\author{C. L\'opez-Sanjuan\inst{1}\fnmsep
\thanks{Based on data obtained
with the European Southern Observatory Very Large Telescope, Paranal,
Chile, under Large Programs 070.A-9007 and 177.A-0837.  Based on
observations obtained with MegaPrime/MegaCam, a joint project of CFHT
and CEA/DAPNIA, at the Canada-France-Hawaii Telescope (CFHT) which is
operated by the National Research Council (NRC) of Canada, the
Institut National des Sciences de l'Univers of the Centre National de
la Recherche Scientifique (CNRS) of France, and the University of
Hawaii. This work is based in part on data products produced at
TERAPIX and the Canadian Astronomy Data Centre as part of the
Canada-France-Hawaii Telescope Legacy Survey, a collaborative project
of NRC and CNRS.}
\and O. Le F\`evre\inst{1} 
\and L. de Ravel\inst{2} 
\and O. Cucciati\inst{1} 
\and O. Ilbert\inst{1} 
\and L. Tresse\inst{1}
\and S. Bardelli  \inst{3}
\and M. Bolzonella  \inst{3} 
\and T. Contini \inst{4}
\and B. Garilli \inst{5}
\and L. Guzzo \inst{6}
\and D. Maccagni \inst{5}
\and H.\ J. McCracken \inst{7,8}
\and Y. Mellier \inst{7,8}
\and A. Pollo \inst{9,10,11}
\and D. Vergani \inst{5}
\and E. Zucca    \inst{3}
}

\institute{Laboratoire d'Astrophysique de Marseille, P\^ole de l'Etoile Site de Ch\^ateau-Gombert 38, rue Fr\'ed\'eric Joliot-Curie, F-13388 Marseille, France \mail{carlos.lopez@oamp.fr}  
\and Institute for Astronomy, University of Edinburgh, Blackford Hill, Edinburgh EH9 3HJ, U.K. 
\and INAF-Osservatorio Astronomico di Bologna, Via Ranzani 1, I-40127, Bologna, Italy 
\and Laboratoire d'Astrophysique de Toulouse-Tarbes, Universit\'e de Toulouse, 
CNRS, 14 av. E. Belin, F-31400 France 
\and IASF-INAF, Via Bassini 15, I-20133, Milano, Italy 
\and INAF-Osservatorio Astronomico di Brera, Via Brera 28, I-20021, Milan, Italy 
\and Institut d'Astrophysique de Paris, UMR 7095, 98 bis Bvd Arago, F-75014, Paris, France 
\and Observatoire de Paris, LERMA, 61 Avenue de l'Observatoire, F-75014, Paris, France 
\and The Andrzej Soltan Institute for Nuclear Studies, ul. Hoza 69, 00-681 
Warszawa, Poland 
\and Astronomical Observatory of the Jagiellonian University, ul Orla 171, PL-30-244, Krak{\'o}w, Poland 
\and Center for Theoretical Physics PAS, Al. Lotnikow 32/46, 02-668 Warsaw, Poland 
} 

\date{}
\date{Received 28 September 2010 / Accepted 17 February 2011}

\abstract
{}
{The role of minor galaxy mergers in galaxy evolution, and in particular to mass assembly, 
is an open question. In this work we measure the merger fraction, $f_{\rm m}$, of $L_{B} \gtrsim L^{*}_{B}$ galaxies in the VVDS-Deep spectroscopic 
survey, and study its dependence on the $B-$band luminosity ratio of the galaxies in the pair, $\mu \equiv L_{B,2}/L_{B,1}$, focusing on minor mergers with $1/10 \leq \mu < 1/4$, and on the rest-frame $NUV-r$ colour of the principal galaxies.}
{We use spectroscopic pairs with redshift $z\lesssim1$ in the VVDS-Deep survey to define kinematical close pairs as those galaxies with a separation on the sky plane $5h^{-1}\ {\rm kpc} < r_{\rm p} \leq r_{\rm p}^{\rm max}$ and a relative velocity $\Delta v \leq 500$ km s$^{-1}$ in redshift space. We vary $r_{\rm p}^{\rm max}$ from $30h^{-1}$ kpc to $100h^{-1}$ kpc. We study $f_{\rm m}$ in two redshift intervals and for several values of $\mu$, from 1/2 to 1/10.}
{The merger fraction dependence on $\mu$ is well described by a 
power-law function, $f_{m}\,(\geq\,\mu) \propto \mu^{s}$. The value of $s$ evolves from $s = -0.60\pm0.08$ at $z = 0.8$ to $s = -1.02\pm0.13$ at $z = 0.5$. The fraction of minor mergers for bright galaxies shows little evolution with redshift as a power-law $(1+z)^m$ with index $m=-0.4\pm0.7$ for the merger fraction and $m=-0.5\pm0.7$ for the merger rate, in contrast with the increase in the major merger fraction ($m = 1.3\pm0.5$) and rate ($m = 1.3\pm0.6$) for the same galaxies. We split our principal galaxies in red and blue, finding that i) $f_{\rm m}$ is higher for red galaxies at every $\mu$, ii) $f_{\rm m}^{\rm red}$ does not evolve with $z$, with $s = -0.79\pm0.12$ at $0.2 < z < 0.95$, and iii) $f_{\rm m}^{\rm blue}$ evolves dramatically: the major merger fraction of blue galaxies decreases by a factor of three with cosmic time, while the minor merger fraction of blue galaxies is roughly constant.}
{Our results show that the mass of normal $L_B \gtrsim L_{B}^{*}$ galaxies has grown by about 25\% since $z\sim1$ because of mergers. The relative contribution of the mass growth by merging is $\sim25$\% due to minor mergers and $\sim$75\% due to major mergers. The relative effect of merging is more important for red than for blue galaxies, with red galaxies subject to 0.5 minor and 0.7 major mergers since $z\sim1$, which leads to a mass growth of $\sim40$\% and a size increase by a factor of 2. Our results also suggest that, for blue galaxies, minor mergers likely lead to early-type spirals rather than elliptical galaxies. These results show that minor merging is a significant but not dominant mechanism contributing to the mass growth of galaxies in the last $\sim 8$ Gyr.}

\keywords{Galaxies : evolution --- galaxies : formation --- galaxies : interactions}

\titlerunning{The minor merger rate of $L_{B} \gtrsim L_{B}^{*}$ galaxies}

\maketitle

\section{Introduction}\label{intro}
As galaxies evolve along cosmic time in the framework of a hierarchical 
assembly of dark matter haloes, a significant fraction of their accreted mass
is expected to come from galaxy-galaxy mergers. The total stellar mass density is 
increasing along cosmic time, faster for early-type galaxies \citep[e.g.,][]{drory05,bundy05,arnouts07,ilbert10}, 
and galaxy-galaxy merging is a natural physical process to participate to this growth.
The role of mergers in galaxy evolution has long been recognised, leading to
mass growth and perturbed morphologies, and mergers have been identified 
as a way to shape elliptical galaxies. 

Major mergers, the encounter of two galaxies of comparable masses leading to a fusion, have
now been well documented in the nearby as well as in the distant universe. 
While the fraction of major mergers in the
nearby Universe is about 2\% \citep{patton00,patton08,darg10i}, it has now been convincingly shown that 
major mergers were more numerous at redshifts up to $z\sim1$ \citep[e.g.,][]{lefevre00,patton02,lin08,deravel09,clsj09ffgoods}, 
with the merger rate of bright/massive galaxies 
staying relatively stable along cosmic time, while the merger rate of intermediate luminosity/mass galaxies 
was stronger in the past \citep{deravel09}. Major mergers have been
shown to contribute a significant but not dominant part of the mass 
growth above the characteristic luminosity $L^*$, with major mergers being
responsible for about 20\% of the stellar mass growth \citep{bundy09,wild09,deravel09,clsj10megoods}. 

As major mergers are apparently not the most important contributor to the mass growth
since $z \sim 1$, other processes need to have taken place. Secular processes such as steady cold accretion
\citep[][and references therein]{genel10} or other mass accretion processes like minor mergers must drive this transformation. 
The merging of smaller galaxies with a more massive one, the minor merger process,
is a possible way to increase the mass of galaxies as minor mergers, if frequent,
could lead to a significant mass increase.
Indirect evidence for minor merging has been presented in the recent literature, including recent star formation in early-type galaxies
being compatible with a minor merger origin \citep{kaviraj07,kaviraj09,onti11}, as confirmed by simulations \citep{mihos94,bournaud07}.

However, so far only few attempts to study the minor merger rate in the local Universe or beyond have been published. 
Unfortunately, to our knowledge, there are no references to the minor merger rate in local galaxies. At higher redshifts,
\citet{lotz08ff} and \citet{jogee09} use distortions in galaxy morphologies to infer that the combined major and minor merger 
fraction is nearly constant since $z \sim 1$. On the other hand, \citet{clsj10pargoods} estimates that 
the major and minor merger rate is $\sim1.7$ times the major rate for $\log\,(M_{\star}/M_{\odot}) \geq 10$ 
galaxies in GOODS-S at $0.2 < z < 1.1$ from their spectro-photometric catalogue. 

Here we report the results from the first measurement of the minor merger fraction and rate using kinematically confirmed close pairs. We use the VVDS-Deep spectroscopic redshift survey which offers a unique combination of deep spectroscopy ($I_{\rm AB} \leq 24$) to identify faint merging companions, and a wide area (0.5 $\deg^{2}$) which contains enough bright galaxies for a statistically robust analysis.

This paper is organized as follows. In Sect.~\ref{data} we summarize the second epoch VVDS-Deep survey data set, while in Sect.~\ref{ncs} the methodology and weight scheme to obtain the merger fraction by close pair statistics and its extension to the regime of minor companions. In Sect.~\ref{ffmu} we measure the merger fraction as a function of the redshift and the luminosity ratio between the galaxies in pairs, while in Sect.~\ref{ffcol} we study the merger fraction of red and blue galaxies. We estimate the minor merger rate of bright galaxies in Sect.~\ref{mrpair}, and we discuss the implications of our results in Sect.~\ref{discussion}. Finally, we present our conclusion in Sect.~\ref{conclusion}. We use $H_0 = 100h\ {\rm km\
s^{-1}\ Mpc^{-1}}$, $h = 0.7$, $\Omega_{M} = 0.3$, and $\Omega_{\Lambda} = 0.7$ throughout. All reported magnitudes are AB.

\section{VVDS-Deep sample}\label{data}

The VVDS-Deep sample\footnote{http://www.oamp.fr/virmos/vvds.htm} \citep{lefevre05} is magnitude selected with 
$17.5 \leq I_{AB} \leq 24$. The spectroscopic survey has been conducted 
on the 0224-04 field with the 
VIMOS multi-slit spectrograph on the VLT \citep{lefevre03}, with 4h integration
using the LRRED grism at a spectral resolution $R\sim230$. 
The multi-slit data processing has been performed using the VIPGI package \citep{scodeggio05}. 
Redshift measurement has followed a strict approach,
with initial guesses based on cross-correlation with reference templates at the 
same redshift, followed by careful eye-checking independently by two team members
before confronting their results. The final redshifts and quality flags follow
a statistically well defined behaviour, leading to a survey for which at least 
80\% of the sample has a secure redshift. This comprises sources with quality
flag = 4 (99\% secure), 3 (95\% secure), 2 (80\% secure) and 9 (those with only a single secure spectral 
feature in emission in their spectrum). The 
accuracy in the redshift measurement is 276 km s$^{-1}$.

Deep photometry is available in this field from a first campaign with
the CFH12K camera (\citealt{lefevre04img} and \citealt{mccracken03}),
followed by very deep observations with the CFHTLS survey. Using 
photometric redshifts \citep{ilbert06phot}, we show that for the
galaxies making up the 20\% incompleteness, about 10\% have a 
tentative (quality flag = 1) spectroscopic redshift which is right for 50\% of them, 
the other 10\% have wrong or unknown spectroscopic redshifts, but we use 
photometric redshift estimates to fully understand the survey completeness as a 
function of magnitude, type, and redshift.

A total of 8359 galaxies with $0 < z_{\rm spec} \leq 1.2$ and $17.5 \leq I_{\rm AB} \leq 24$ 
(primary objects with flags = 1,2,3,4,9; and secondary objects, those that lie by chance in the slits, with
flags = 21, 22, 23, 24, 29) from second epoch VVDS-Deep data (Le F\`evre et al., in prep.) have been used in this paper. 
Note that we have used flag = 1 sources, which are 50\% secure and 
that have not been used in previous VVDS-Deep works, thanks
to the improved weighting scheme in VVDS-Deep (see Sect.~\ref{complet}, for details). 
 
\section{Statistics of minor close companions in spectroscopic samples}\label{ncs}

In this section we review the commonly used methodology for computing major merger fractions by close pair statistics in spectroscopic samples, and we extend it to search for minor (i.e., faint) companions in the VVDS-Deep.

The distance between two sources can be measured as a function of their projected separation, 
$r_{\rm p} = \theta d_A(z_i)$, and their rest-frame relative velocity along the line of sight, 
$\Delta v = {c|z_j - z_i|}/(1+z_i)$, where $z_i$ and $z_j$ are the redshift of the principal (more luminous galaxy in the pair) and the companion galaxy, respectively; $\theta$ is the angular separation, 
in arcsec, of the two galaxies on the sky plane; and $d_A(z)$ is the angular scale, in kpc/arcsec, at redshift $z$. 
Two galaxies are defined as a close pair if $r_{\rm p}^{\rm min} < r_{\rm p} \leq r_{\rm p}^{\rm max}$ and 
$\Delta v \leq \Delta v^{\rm max}$. The inner limit in $r_{\rm p}$ is imposed to avoid spatial resolution limitations
due to the size of the observed point spread function. 
Reasonable limits for ground-based data are $r_{\rm p}^{\rm min} = 5h^{-1}$ kpc, $r_{\rm p}^{\rm max} = 20h^{-1}$ kpc, and 
$\Delta v^{\rm max} = 500$ km s$^{-1}$. With these constraints, it is expected that 
50\%-70\% of the selected close pairs will finally merge \citep{patton00,patton08,lin04,lin10,bell06}. 
We used $\Delta v^{\rm max} = 500\ {\rm km\, s^{-1}}$, $r_{\rm p}^{\rm min} = 5h^{-1}$ kpc, and varied the 
value of $r_{\rm p}^{\rm max}$ from $30h^{-1}$ kpc to $100h^{-1}$ kpc to study the dependence of the merger fraction 
with the surrounding volume. 

We select principal galaxies as defined below and we look for companion galaxies that fulfill the close pair criterion for each galaxy of the principal sample. If one principal galaxy has more than one close companion, we take each possible pair separately (i.e., for the close galaxies A,B, and C, we study the pairs A-B, B-C, and A-C as independent). In addition, we impose a rest-frame $B$-band luminosity difference between the pair members. We denote the ratio between the luminosity of the principal galaxy, $L_{B,1}$, and the companion galaxy, $L_{B,2}$, as 
\begin{equation}
\mu \equiv \frac{L_{B,2}}{L_{B,1}},
\end{equation}
and looked for those systems with $L_{B,2} \geq \mu L_{B,1}$ or, equivalently, $M_{B,2} - M_{B,1} \leq \Delta M_{B} = -2.5\log \mu$, where $M_{B,1}$ and $M_{B,2}$ are the $B-$band absolute magnitudes of the principal and companion galaxy in the pair, respectively. We define as major companions those close pairs with $\mu \geq 1/4$, while minor companions those with $1/10 \leq \mu < 1/4$.

\begin{figure}[t!]
\resizebox{\hsize}{!}{\includegraphics{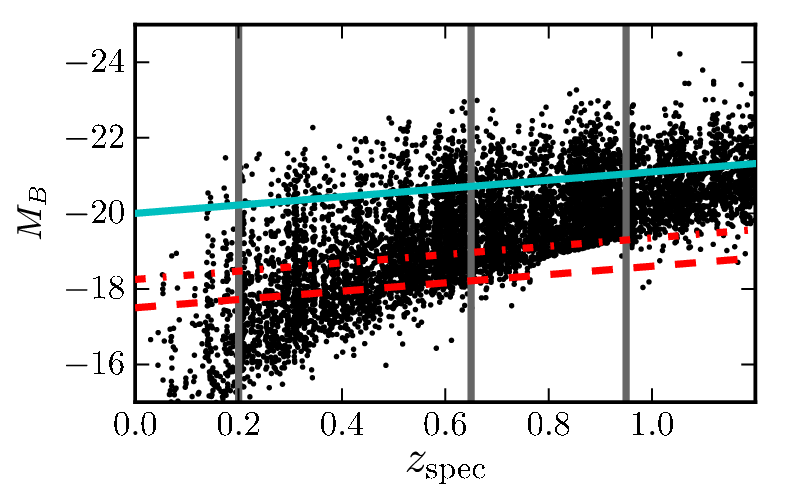}}
\caption{$B-$band absolute magnitude versus redshift for all the VVDS-Deep sources with $z_{\rm spec} \leq 1.2$. Vertical solid lines identify the redshift intervals in our study, named $z_{\rm r,1} = [0.2,0.65)$ and $z_{\rm r,2} = [0.65,0.95)$. The horizontal solid line represents the selection of the principal galaxies sample, $M_{B,1} \leq -20 - 1.1z$. The dashed line shows the limit of the companion sample down to $\mu \geq 1/10$, $M_{B,2} \leq -17.5 - 1.1z$, while the dash--dotted line shows that we are complete in both redshift bins when we search for $\mu \geq 1/5$ companions, $M_{B,2} \leq -18.25 - 1.1z$. [{\it A colour version of this plot is available at the electronic edition}].}
\label{mbzfig}
\end{figure}

We aimed to reach the minor companion regime, i.e., $\mu = 1/10$ ($\Delta M_{B} = 2.5$). 
For this, we define our principal galaxy sample and companions, and redshift ranges, 
to preserve statistical robustness and to minimize 
completeness corrections (see next section). We select as principal galaxies those with 
$M_B^{\rm e} \leq -20 \sim M_{B}^{*}$ \citep[e.g.,][]{ilbert05}, where $M_B^{\rm e} = M_B + Qz$ 
and the constant $Q = 1.1$ accounts for the evolution of the luminosity function in VVDS-Deep survey \citep{ilbert05}.
With this limit, companions with $\mu$ down to 1/10 will be included in the VVDS-Deep sample (Fig.~\ref{mbzfig}).
Thanks to the wide area of VVDS-Deep, we have 1011 principal galaxies at $0.1 < z < 1.0$. To study minor companions we define as 
companion galaxies those with $M_B^{\rm e} \leq -17.5$, and impose different luminosity ratios, 
$\mu \geq 1/2$, 1/3, 1/4, 1/5, 1/6, 1/7, 1/8, and 1/10 ($\Delta M_{B} = 0.75$, 1.2, 1.5, 1.75, 1.95, 2.1, 2.25, and 2.5, 
respectively). We define two redshift bins, named $z_{\rm r,1} = [0.2,0.65)$ and $z_{\rm r,2} = [0.65,0.95)$. In these bins, the mean redshifts of the principal galaxies, weighted to take into account their spectroscopic completeness (see next section for details), are $\overline{z_{\rm r,1}} = 0.5$ and $\overline{z_{\rm r,2}} = 0.8$. In the former we are complete for $\mu \geq 1/10$ companions, while in the latter we reach $\mu \geq 1/5$ (Fig.~\ref{mbzfig}), 
therefore requiring a completeness correction for $1/10 \leq \mu < 1/5$ companions (Sect.~\ref{complet}). We are able to reach this faint companions regime due to the depth of the VVDS-Deep spectroscopy ($I_{\rm AB} \leq 24$). The number of principal galaxies is $n_{1} = 351$ at $z_{\rm r,1}$ and $n_{2} = 544$ at $z_{\rm r,2}$, this is, $n_2/n_1$ = 1.55. On the other hand, the ratio between the probed cosmological volumes is $V_2/V_1 = 1.52$, so the number density of principal galaxies is similar in both ranges. Using the group catalog from the VVDS-Deep second-epoch data presented in \citet{cucciati10}\footnote{We did not use galaxies with flag = 1 and 21 in groups determination. However, only 2\% of the principal galaxies have flag = 1 or 21 because they are bright.} we find that 14\%/13\% of principal galaxies at $z_{\rm r,1}$/$z_{\rm r,2}$ are in a group with three or more members. Hence, also the environment of our principal galaxies is similar in both ranges under study.

If we find $N_{\rm p}$ close pairs in our sample for a given luminosity ratio $\mu$, the merger fraction is
\begin{equation}
f_{m}\,(\geq \mu) = \frac{N_{\rm p}\,(\geq\,\mu)}{N_{1}},
\end{equation}
where $N_{1}$ is the number of galaxies in the principal sample.  With this definition the merger fraction is cumulative when $\mu$ decrease. This simple definition is valid for volume-limited samples, while we work with spectroscopic, luminosity-limited samples. Because of this, we must take into account the different selection effects in our measurement of the merger fraction. 

\subsection{Accounting for selection effects}\label{complet}
Following \citet{deravel09}, we correct for three basic selection effects:
\begin{enumerate}
\item the limiting magnitude $I_{\rm AB} = 24$ which imposes a loss of faint companions.
\item the spatial sampling rate and the spectroscopic success rate in measuring redshifts.
\item the loss of pairs at small separations because of the ground based seeing limitation of the observations.
\end{enumerate}

The spectroscopic targets have been selected on the basis of the magnitude criterion $17.5 \leq I_{\rm AB} \leq 24$. Therefore, we miss companions of the principal galaxies which have an absolute magnitude fainter than imposed by the $I_{\rm AB} = 24$ cut off and the $\Delta M_{B}$ magnitude difference, artificially lowering the number of pairs. To take this into account we compute for each observed pair a weight $w^{k}_{\rm mag}(M_{B,1},z)$ using the ratio between the co-moving number densities above and below the magnitude cut off \citep{patton00}:
\begin{equation}
w^k_{\rm mag} (M_{B,1},z)= \frac{\int^{M_{B,{\rm sup}}^k}_{-\infty} \Phi(M_{B},z)\,{\rm d}M_B} {\int^{M_{B,{\rm lim}}(z)}_{-\infty} \Phi(M_{B},z)\,{\rm d}M_B},\label{snz}
\end{equation}
where $M_{B,{\rm lim}}(z)$ is the limiting magnitude of the catalogue at redshift $z$, $M_{B,{\rm sup}}^k = M_{B,1}^{k} + \Delta M_{B}$ is the lower luminosity of a close companion of the principal galaxy in the pair $k$, and $\Phi(M_{B},z)$ is the luminosity function in the $B$-band at redshift $z$. We assumed the luminosity function measured in the VVDS-Deep area by \citet[][see also \citealt{zucca06}]{ilbert05}. We take $w^k_{\rm mag} = 1$ when $M_{B,{\rm sup}}^k \leq M_{B,{\rm lim}}(z)$. We note that the number of companions with $\mu \geq 1/10$ is complete for all principal sources with $M_B^{\rm e} \leq -20$ at $z_{\rm r, 1}$ and $\sim50$\% at $z_{\rm r,2}$ (Fig.~\ref{mbzfig}), while the completeness is $\sim70$\%, 80\%, 90\% and 100\% at $z_{\rm r,2}$ for companions with $\mu \geq 1/8, 1/7, 1/6$ and $1/5$, respectively. That is, $w^k_{\rm mag} \neq 1$ only for $1/10 \leq \mu < 1/5$ companions of some systems at $0.65 \leq z < 0.95$. We further test the weights $w^k_{\rm mag}$ in Sect.~\ref{ffmu}.

Since $\sim$25\% of the total number of potential targets in the VVDS-Deep field have been spectroscopically observed and the redshifts are not measured with 100\% certainty, we must correct for the VVDS-Deep target sampling rate and redshift success rate. These have been well constrained resulting in the Target Sampling Rate (TSR) and the Spectroscopic Success Rate (SSR) computed as a function of redshift, source magnitude and source size ($x$). The SSR has been assumed independent of the galaxy type, as demonstrated up to $z \sim 1$ in \citet{zucca06}. As several first epoch VVDS-Deep galaxies with flag 1 and 2 have been re-observed in the VVDS-Ultradeep survey ($I_{\rm AB} \leq 24.75$, Le F\`evre et al., in prep.), providing a robust measurement of their redshift, this offers the opportunity to estimate the reliability of VVDS-Deep flag = 1 and 2 sources, and we define a weight $w_{129}$ to take this into account. We also define the weight $w_{129}$ for flag = 9 sources by comparison with the latest photometric redshifts in the VVDS-Deep field \citep[see][for details about the latest photometric data set in this field]{cucciati10}. By definition, $w_{129} = 1$ for flag = 3 and 4 sources. We derived the spectroscopic completeness weight for each galaxy $i$ in the catalogue as
\begin{equation}
w^{i}_{\rm spec}(z,I_{\rm AB},x) = \frac{1}{TSR^{i} \times SSR^{i} \times w_{129}^{i}},
\end{equation}
and assigned a weight $w^{k}_{\rm spec} =  w^{1}_{\rm spec} \times w^{2}_{\rm spec}$ at each close pair, where $w^{1}_{\rm spec}$ and $w^{2}_{\rm spec}$ are the spectroscopic completeness weights of the principal and the companion galaxy in the pair, respectively.

The last correction we need to apply results from the observations which have been performed under a typical ground based seeing of $1^{\prime\prime}$. We correct for the increasing incompleteness in targeting both components of close pairs as the separation between them is getting smaller. Assuming a clustered distribution of galaxies, the number of galaxy pairs should be a monotonically decreasing function of the pair separation \citep[e.g.,][]{bell06,lin08}. However, pairs start to be under-counted for separations $\theta  \leq 2^{\prime\prime}$ because of seeing effects. We apply a weight $w_{\theta}^{k}$ on each pair using the ratio
\begin{equation}
w_{\theta}^{k} = \frac{a}{r_{\rm zz}\,(\theta_k)},
\end{equation}
where the mean ratio $a$ is the probability to randomly select a pair, obtained at large separations, and 
$r_{\rm zz}\,(\theta_k)$ is the ratio between the observed pair count in the spectroscopic catalogue, $N_{\rm zz}$,  
over the observed pair count in the photometric one, $N_{\rm pp}$. For large separations ($\theta > 50^{\prime\prime}$), 
$r_{\rm zz} \sim a$, but at small separations $r_{\rm zz} < a$ because of the artificial decrease of pairs due to seeing effects (see \citealt{deravel09}, for further details). This weight also accounts for other geometrical biases in the survey, e.g., those related with the minimum separation between slits. Compared to the weight 
$w_{\theta}^{k}$ for the total major merger population  \citep{deravel09}, the weight for faint companions 
could be different as it is  more difficult to measure the $z_{\rm spec}$ of  fainter galaxies 
located near a bright principal galaxy. 
To explore this possibility, we compare the number of 
photometric and spectroscopic pairs for a given angular distance and luminosity difference in the $I_{\rm AB}$ 
band between the pair members ($\Delta I_{\rm AB}$). We study the variation of $r_{\rm zz}\,(\theta_k,\Delta I_{\rm AB})$ 
from $\theta = 1^{\prime\prime}$ to $100^{\prime\prime}$ for four different luminosity differences, 
$\Delta I_{\rm AB} \leq 0.75$, $0.75 < \Delta I_{\rm AB} \leq 1.5$, $1.5 < \Delta I_{\rm AB} \leq 2$, and 
$2 < \Delta I_{\rm AB} \leq 2.5$. We find that in all cases $r_{\rm zz}$ tends to become constant at large angular 
separations, while at $\theta \lesssim 10^{\prime\prime}$ the value of $r_{\rm zz}$ tends to be lower for higher 
$\Delta I_{\rm AB}$, making it more difficult to recover a faint companion than a bright one. However, 
when compared with the global value of $r_{\rm zz}$, this systematic effect leads to differences $\lesssim5$\%. 
Because the dispersion in the global $w_{\theta}^{k}$ is $\sim10$\%, we have decided not to apply any correction 
to this systematic effect.

Finally, the corrected merger fraction is 
\begin{equation}
f_{\rm m}\,(\geq\,\mu) = \frac{\sum^{N_{\rm p}\,(\geq\,\mu)}_{k}\,w_{\rm spec}^{k} w_{\rm mag}^{k} w_{\theta}^{k}}{\sum^{N_{1}}_{i} w_{\rm spec}^{i}}.\label{fms}
\end{equation}

In order to estimate the error of $f_{\rm m}$ we used the jackknife technique \citep{efron82}. We computed partial standard deviations, $\delta_k$, for each system $k$ by taking the difference between the measured $f_{\rm m}$ and the same quantity with the $k$th pair removed for the sample, $f_{\rm m}^k$, such that $\delta_k = f_{\rm m} - f_{\rm m}^k$. For a sample with $N_{\rm p}$ systems, the variance is given by $\sigma_{f_{\rm m}}^2 = [(N_{\rm p}-1) \sum_k \delta_k^2]/N_{\rm p}$. We checked that the variances estimated by jackknife technique are similar, within $\sim10$\%, to those estimated by a Bayesian approach \citep{cameron10sig}.

\begin{figure}[t!]
\resizebox{\hsize}{!}{\includegraphics{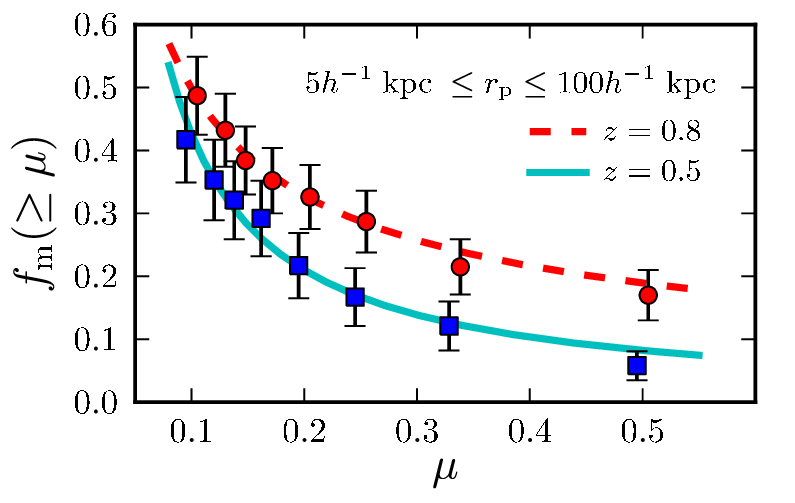}}
\caption{Merger fraction versus luminosity ratio in $B-$band, $\mu$, for close pairs with $r_{\rm p}^{\rm min} = 5h^{-1}$ kpc and $r_{\rm p}^{\rm max} = 100h^{-1}$ kpc. Dots are the merger fractions at $z = 0.8$, and squares at $z = 0.5$. The lines are the GLS fits of a power-law, $f_{\rm m}\,(\geq\,\mu) \propto \mu^{s}$, to the $z = 0.8$ ($s = -0.60$; dashed) and $z = 0.5$ data ($s = -1.02$; solid). [{\it A colour version of this plot is available at the electronic edition}].}
\label{ffmufig}
\end{figure}

\begin{table}
\caption{Merger fraction of $L_{B,1} \gtrsim L_{B}^{*}$ galaxies for $r_{\rm p}^{\rm max} = 100h^{-1}$ kpc as a 
function of luminosity ratio $\mu$.}
\label{ffmutab}
\begin{center}
\begin{tabular}{lccccc}
\hline\hline\noalign{\smallskip}
$\mu$ & \multicolumn{2}{c}{$z = 0.5$} & & \multicolumn{2}{c}{$z = 0.8$}\\
\noalign{\smallskip}
\cline{2-3} \cline{5-6}
\noalign{\smallskip}
      & $N_{\rm p}\,(\geq\,\mu)$ & $f_{\rm m}\,(\geq\,\mu)$ & & $N_{\rm p}\,(\geq\,\mu)$ & $f_{\rm m}\,(\geq\,\mu)$\\
\noalign{\smallskip}
\hline
\noalign{\smallskip}
1/2  &  6 & $0.058 \pm 0.023$ & & 22 & $0.169 \pm 0.040$ \\
1/3  & 11 & $0.121 \pm 0.039$ & & 29 & $0.215 \pm 0.043$ \\
1/4  & 15 & $0.167 \pm 0.046$ & & 39 & $0.287 \pm 0.049$ \\
1/5  & 20 & $0.216 \pm 0.051$ & & 45 & $0.322 \pm 0.050$ \\
1/6  & 26 & $0.291 \pm 0.060$ & & 49 & $0.347 \pm 0.052$ \\
1/7  & 29 & $0.320 \pm 0.062$ & & 53 & $0.379 \pm 0.054$ \\
1/8  & 33 & $0.351 \pm 0.064$ & & 58 & $0.426 \pm 0.057$ \\
1/10 & 40 & $0.413 \pm 0.067$ & & 63 & $0.479 \pm 0.061$ \\
\hline
\end{tabular}
\end{center}
\end{table}

\section{The minor merger fraction of $L_{B} \gtrsim L_{B}^{*}$ galaxies}\label{ffmu}
In this section we study the merger fraction of bright galaxies as a function of $\mu$, reaching the minor companion regime ($1/10 \leq \mu < 1/4$) with spectroscopically confirmed close pairs. We summarize the values of $f_{\rm m}\,(\geq\,\mu)$ obtained at $z_{\rm r, 1} = [0.2,0.65)$ and $z_{\rm r, 2} = [0.65,0.95)$ for $r_{\rm p}^{\rm max} = 100h^{-1}$ kpc and different luminosity ratios in Table~\ref{ffmutab}, and show them in Fig.~\ref{ffmufig}. The merger fraction decreases with cosmic time for all $\mu$, but this difference is lower for smaller $\mu$ values. The merger fraction at both redshift bins increases when $\mu$ decreases, a natural consequence of our $f_{\rm m}\,(\geq \mu)$ definition as the fraction of principal galaxies with a $L_{B,2} \geq \mu L_{B,1}$ companion.

\begin{figure}[t!]
\resizebox{\hsize}{!}{\includegraphics{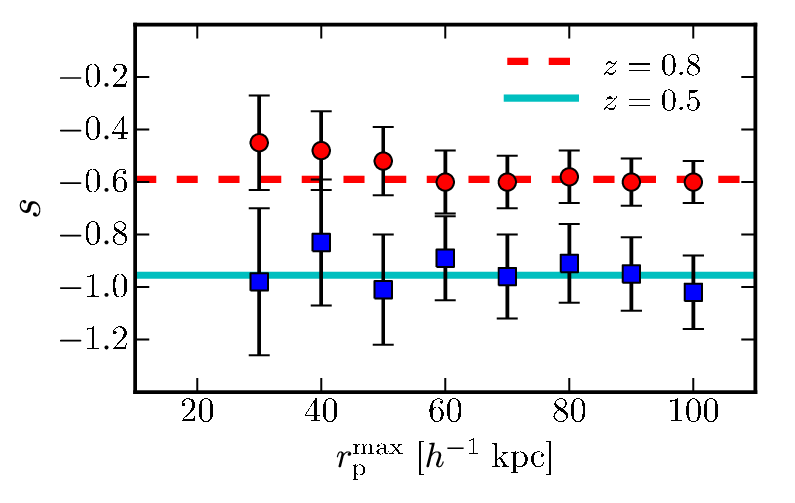}}
\caption{Power-law index $s$ versus $r_{\rm p}^{\rm max}$. Dots are for $z = 0.8$ galaxies, and squares for $z = 0.5$ galaxies. The lines are the median of the data: $s = -0.59$ at $z = 0.8$ (dashed) and $s = -0.96$ at $z = 0.5$ (solid). [{\it A colour version of this plot is available at the electronic edition}].}
\label{svszfig}
\end{figure}

The observed dependence of $f_{\rm m}$ on $\mu$ is well parametrized as 
\begin{equation}
f_{\rm m}\,(\geq \mu) = f_{\rm MM}\,\bigg(\frac{\mu}{\mu_{\rm MM}}\bigg)^{s},\label{ffmueq}
\end{equation}
where $f_{\rm MM}$ is the major merger fraction ($\mu \geq \mu_{\rm MM} = 1/4$). This dependence was predicted by 
the cosmological simulations of \citet{maller06} and used by \citet{clsj10pargoods} in mass-selected 
spectro-photometric close pairs. We set the value of $f_{\rm MM}$ to the observed one and used Generalized Least Squares (GLS) to estimate the power-law index $s$ (see Appendix~\ref{mcfit}, for details). The GLS fit to the Table~\ref{ffmutab} data 
yields $s = -0.60\pm 0.08$ at $z = 0.8$ and $s = -1.02 \pm 0.13$ at $z = 0.5$. To obtain a robust value of $s$ 
at each redshift range under study, we determine $s$ for different $r_{\rm p}^{\rm max}$. We summarize our results 
in Table~\ref{srptab} and show them in Fig.~\ref{svszfig}. The values of $s$ measured at $r_{\rm p}^{\rm max} = 100h^{-1}$ 
are representative of the median of all the values at different $r_{\rm p}^{\rm max}$, that are 
$s = -0.59$ at $z = 0.8$ and $s = -0.96$ at $z = 0.5$.

\begin{table}
\caption{Power-law index $s$ as a function of search radius $r_{\rm p}^{\rm max}$}\label{srptab}
\begin{center}
\begin{tabular}{lcc}
\hline\hline\noalign{\smallskip}
$r_{\rm p}^{\rm max}$ & $z = 0.5$ & $z = 0.8$\\
($h^{-1}$ kpc)& &\\
\noalign{\smallskip}
\hline
\noalign{\smallskip}\vspace{0.5mm}
30  & $-0.98 \pm 0.28$ & $-0.45 \pm 0.18$ \\
40  & $-0.83 \pm 0.24$ & $-0.48 \pm 0.15$ \\
50  & $-1.01 \pm 0.21$ & $-0.52 \pm 0.13$ \\
60  & $-0.89 \pm 0.16$ & $-0.60 \pm 0.12$ \\
70  & $-0.96 \pm 0.16$ & $-0.60 \pm 0.10$ \\
80  & $-0.91 \pm 0.15$ & $-0.58 \pm 0.10$ \\
90  & $-0.95 \pm 0.14$ & $-0.60 \pm 0.09$ \\
100 & $-1.02 \pm 0.14$ & $-0.60 \pm 0.08$ \\
\hline
\end{tabular}
\end{center}
\end{table}

We find that the value of $s$ decreases with 
cosmic time, reflecting a differential evolution in the merger fraction of major and minor companions. We checked that 
our incompleteness in the range $z_{\rm r,2}$ (Sect.~\ref{ncs}) does not bias our results with the following test. 
We define a companion sample with $M_B \leq -17.17 - 2.8z$. This sample becomes artificially incomplete for companions with $\mu \geq 1/10$ 
and $\mu \geq 1/5$ at $z \geq 0.2$ and $z \geq 0.65$, respectively; that is, in our first redshift bin, and mimic 
the completeness behaviour of our companion sample at $z_{\rm r,2}$. Then, we repeat the previous analysis with the artificially incomplete sample, 
obtaining $s = -0.99\pm0.08$, which is similar to the original value measured in the complete sample. 
This implies that the weights $w_{\rm mag}^{k}$ properly account for the missing faint companions and that the observed 
evolution of the index $s$ with redshift in VVDS-Deep is a robust result. We also study how the luminosity function assumed in $w_{\rm mag}^{k}$ determination affects the measured merger fractions. We used the $B-$band luminosity functions from \citet{giallongo05,faber07}; and \citet{zucca09}, finding a variation lower than $3$\% in the values of the merger fraction for every $r_{\rm p}^{\rm max}$ compared to our results. Hence, assuming a different luminosity function would have only a limited impact on our results.

\begin{figure}[t!]
\resizebox{\hsize}{!}{\includegraphics{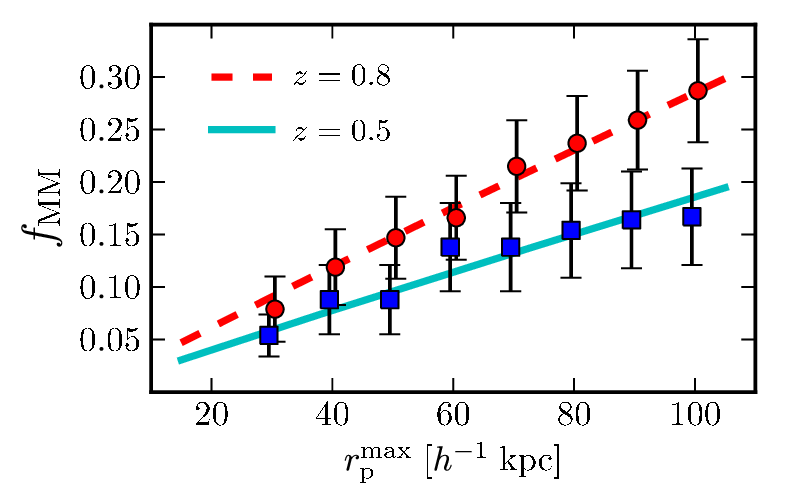}}
\caption{Major merger fraction, $f_{\rm MM}$, versus $r_{\rm p}^{\rm max}$. Dots are for $z = 0.8$ galaxies, and squares for $z = 0.5$ galaxies. The lines are the least-squares best fit of a power-law function, $f_{\rm MM}\,\propto r_{\rm p}^{q}$, to the data. In both cases the power-law index is $q = 0.95$. The points are shifted to avoid overlap. [{\it A colour version of this plot is available at the electronic edition}].}
\label{brpfig}
\end{figure}

We then studied the dependency of the major merger fraction, $f_{\rm MM}$, on the search radius. We summarize the $f_{\rm MM}$ values for all $r_{\rm p}^{\rm max}$ under study in Table~\ref{f1tab} and show them in Fig.~\ref{brpfig}. The value of $f_{\rm MM}$ increases with the search radius  and is well described  in both redshift ranges by a power-law with index $q = 0.95\pm0.20$. Regarding redshift evolution, the major merger fraction increases with redshift, in agreement with previous results in the literature \citep[e.g.,][]{lefevre00,conselice06ff,rawat08,deravel09,clsj09ffgs,clsj09ffgoods}. We study this evolution in more details in Sect.~\ref{ffmmevol}.

\begin{table}
\caption{Major merger fraction of $L_{B,1} \gtrsim L_{B}^{*}$ galaxies, $f_{\rm MM}$, as a function of search radius $r_{\rm p}^{\rm max}$}
\label{f1tab}
\begin{center}
\begin{tabular}{lccc}
\hline\hline\noalign{\smallskip}
$r_{\rm p}^{\rm max}$ & $z = 0.5$ & $z = 0.8$ \\
($h^{-1}$ kpc) & & \\
\noalign{\smallskip}
\hline
\noalign{\smallskip}
30 	& $0.054\pm0.020$	& $0.079\pm0.031$ \\
40	& $0.088\pm0.033$	& $0.110\pm0.036$ \\
50 	& $0.088\pm0.033$ 	& $0.147\pm0.039$ \\
60 	& $0.138\pm0.042$ 	& $0.166\pm0.040$ \\
70 	& $0.138\pm0.042$ 	& $0.215\pm0.044$ \\
80 	& $0.154\pm0.045$ 	& $0.237\pm0.045$ \\
90 	& $0.164\pm0.046$ 	& $0.259\pm0.047$ \\
100  	& $0.167\pm0.046$ 	& $0.287\pm0.049$ \\
\hline
\end{tabular}
\end{center}
\end{table}

\begin{figure}[t]
\centering
\resizebox{\hsize}{!}{\includegraphics{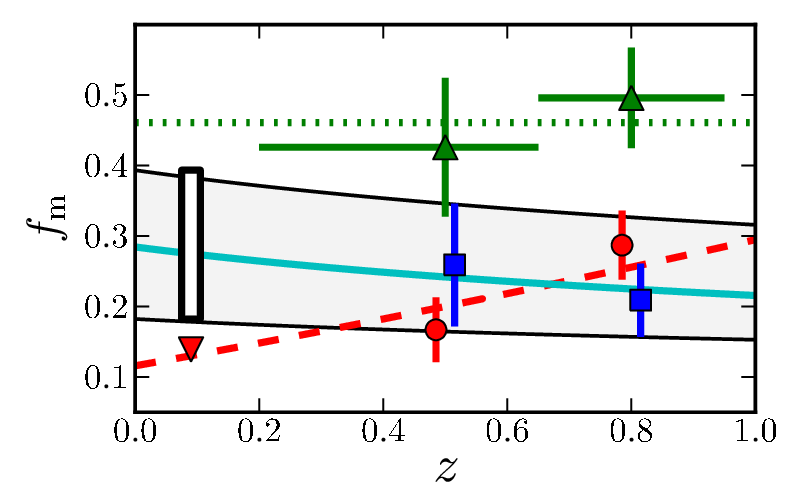}}
\caption{Minor (squares), major (dots), and major + minor (triangles) merger fraction of $M_B^{\rm e} \leq -20$ galaxies for 
$r_{\rm p}^{\rm max} = 100h^{-1}$ kpc as a function of redshift. The points are shifted when necessary to avoid overlap. The $z$ error bars in the total merger fraction mark the redshift range spanned by VVDS-Deep data. The inverted triangle is the major merger 
fraction at $z = 0.09$ from MGC. The white rectangle is the local ($z = 0.09$) minor merger fraction derived from the total and the major merger ones, while the gray area identifies the most probable minor merger fraction values in the range $0 < z < 1$ (see text for details). The solid line is the best fit of a power-law function with a fixed index, $f_{\rm mm} \propto (1+z)^{-0.4}$, to the minor merger fraction data. The dashed line 
is the least-squares best fit of a power-law function to the major merger fraction data. The dotted line is the major + minor merger fraction if it is assumed constant. [{\it A colour version of this plot is available at the electronic edition}].}
\label{ffmmfig}
\end{figure}

We can estimate the minor-to-major merger fraction ratio, denoted $f_{m/M}$, as
\begin{equation}
f_{m/M} \equiv \frac{f_{\rm mm}}{f_{\rm MM}} = \frac{f_{\rm m}\,(\mu_{\rm mm} \leq \mu < \mu_{\rm MM})}{f_{\rm m}\,(\mu \geq \mu_{\rm MM})} = \bigg(\frac{\mu_{\rm mm}}{\mu_{\rm MM}}\bigg)^{s} - 1,
\end{equation}
where $\mu_{\rm MM}$ and $\mu_{\rm mm}$ are the luminosity ratios for major and minor mergers, respectively. 
This definition does not depend on the normalization of the merger fraction, that varies with $r_{\rm p}^{\rm max}$ 
(Fig.~\ref{brpfig}). We assume $\mu_{\rm MM} = 1/4$ and $\mu_{\rm mm} = 1/10$. We find that $f_{m/M} = 0.73 \pm 0.13$ at $z = 0.8$, and $f_{m/M} = 1.55\pm0.30$ at $z = 0.5$. Therefore, minor companions become more numerous than major ones 
as one is going to lower redshifts. To illustrate this, and to facilitate future comparisons, we summarize our best 
estimation of the minor merger fraction for $r_{\rm p}^{\rm max} = 30h^{-1}, 50h^{-1}$, and $100h^{-1}$ kpc in Table~\ref{ffmmtab}, 
and show the minor, major and total (major + minor) merger fractions for $r_{\rm p}^{\rm max} = 100h^{-1}$ kpc in Fig.~\ref{ffmmfig}. The typical error 
in the minor merger fraction is $\sim30-40$\%. Our measurements seem to indicate that the minor merger fraction increases with 
cosmic time. This trend becomes more robust when we further compare our results to a local ($z \sim 0.1$) estimation of the minor merger fraction, Sect.~\ref{ffmmevol}.

\begin{table}
\caption{Minor merger fraction, $f_{\rm m}\,(1/10 \leq \mu < 1/4)$, of $L_{B,1} \gtrsim L^{*}_{B}$ galaxies}
\label{ffmmtab}
\begin{center}
\begin{tabular}{lcc}
\hline\hline\noalign{\smallskip}
$r_{\rm p}^{\rm max}$ & $z = 0.5$ & $z = 0.8$ \\
($h^{-1}$ kpc) & &  \\
\noalign{\smallskip}
\hline
\noalign{\smallskip}
30  & $0.084\pm0.035$ & $0.058\pm0.025$ \\
50  & $0.136\pm0.057$ & $0.107\pm0.034$ \\
100 & $0.259\pm0.087$ & $0.209\pm0.052$ \\
\hline
\end{tabular}
\end{center}
\end{table}

\section{The minor merger fraction of red and blue galaxies}\label{ffcol}
In this section we study the merger fraction as a function of the blue or red colour of the principal galaxy in the pair. 
To split our $M_B^{\rm e} \leq -20$ galaxies into red and blue, we study their distribution in the $M_{NUV} - M_{r}$ 
versus $M_{r} - M_{J}$ plane. The UV -- optical colours is a better tracer of recent star formation than typical 
optical -- optical colours \citep{wyder07,schi07,arnouts07,kaviraj07}, while the addition of an optical -- infrared 
colour to the UV -- optical helps to break the degeneracy between old and dusty star-forming (SF) red galaxies 
\citep{williams09,ilbert10}. Another possibility to separate old and dusty red galaxies is to perform a dust reddening 
correction. This also makes possible a clean separation between the red quiescent sequence and the blue star-forming 
cloud, since the "green valley" region between both sequences is mainly populated by dusty SF galaxies 
\citep{wyder07,cortese08,salim09,brammer09}.

\begin{figure}[t]
\centering
\includegraphics[width = 8.5cm]{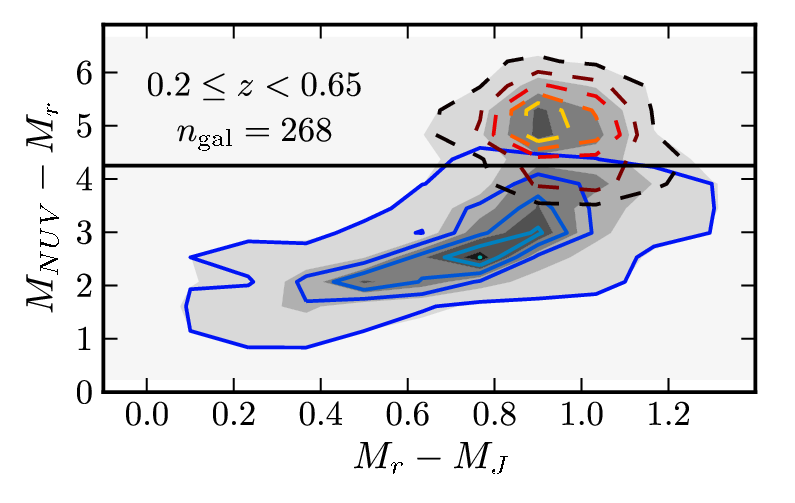}
\includegraphics[width = 8.5cm]{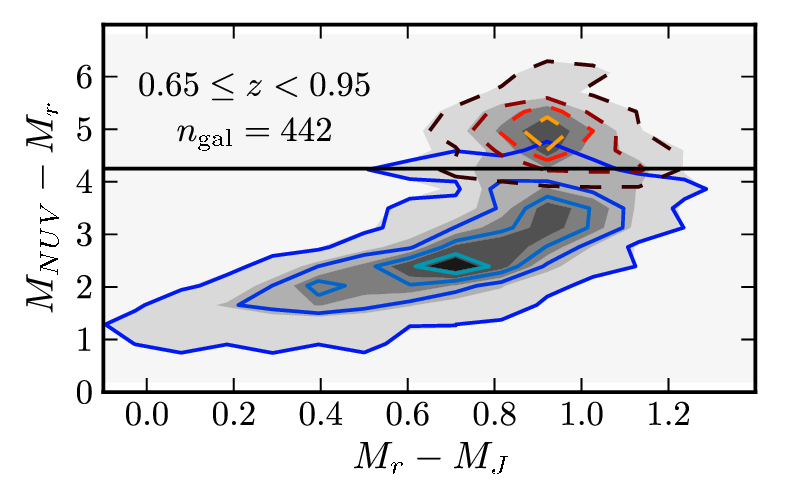}
\caption{Number density (gray scales) of $M_B^{\rm e} \leq -20$ galaxies in the $M_{NUV}-M_{r}$ versus $M_{r} - M_{J}$ plane at $z_{\rm r,1} = [0.2,0.65)$ (top panel) and $z_{\rm r,2} = [0.65,0.95)$ (bottom panel). Dashed and solid contours are the number density of early ($S_{\rm type} \leq 8$) and late ($S_{\rm type} > 8$) spectro-photometric types, respectively. We show those galaxies detected in the $K$ band. The number of sources in each interval, $n_{\rm gal}$, is labeled in the panels. The black solid line is the condition $M_{NUV} - M_{r} = 4.25$ that we use to split our galaxies into red and blue. [{\it A colour version of this plot is available at the electronic edition}].}
\label{nuvrjtotfig}
\end{figure}

In Fig.~\ref{nuvrjtotfig}, we show the number density contours of $M_B^{\rm e} \leq -20$ galaxies in the $M_{NUV} - M_{r}$ versus $M_{r} - M_{J}$ plane for the two redshifts ranges under study, $z_{\rm r,1} = [0.2,0.65)$ and $z_{\rm r,2} = [0.65,0.95)$. We only show those galaxies detected in the $K$ band to avoid that $M_{J}$ was an extrapolation from the fit to the optical photometry. We find a red sequence and a blue cloud in both redshift ranges, as expected from previous works \citep[e.g.,][]{arnouts07,franzetti07}. Both populations are well separated using a constant cut $M_{NUV} - M_{r} = 4.25$. Because of our rest-frame $B-$band luminosity selection, we do not find a significant population of red ($M_{NUV} - M_{r} \gtrsim 4$), dusty SF ($M_{r} - M_{J} \gtrsim 1$) galaxies (i.e., they are faint due to the dust extinction). In contrast, this population appears in NIR-selected samples, as those from \cite{ilbert10} or \cite{bundy10}. To explore in more details the nature of red and blue sources, we use the spectro-photometric types ($S_{\rm types}$) of the galaxies. These spectro-photometric types were obtained by fitting 62 templates, that include ellipticals and S0's ($S_{\rm type} = 1-13$), early-type spirals ($S_{\rm type} = 14-29$), late-type spirals ($S_{\rm type} = 30-43$), and irregulars and starburst ($S_{\rm type} = 44-62$; see \citealt{zucca06}, for details). In Fig.~\ref{nuvrjtotfig}, we also show the number density contours of $M_B^{\rm e} \leq -20$ galaxies when we split them into early ($S_{\rm types} \leq 8$) and late ($S_{\rm types} > 8$) types. We show that, as expected, red sequence galaxies are mainly ($\sim 90$\%) early types, while blue cloud is populated ($\sim 95$\%) by later types \citep[see also][]{arnouts07}. Because of this, and for simplicity, we define red, quiescent galaxies as those with $M_{NUV} - M_{r} \geq 4.25$, and blue, star-forming galaxies as those with $M_{NUV} - M_{r} < 4.25$. We note that the trends and main results in this section remain the same if we either vary the blue--red limit by $\pm0.25$ mag or use spectro-photometric types to define an early (i.e., red) and a late (i.e., blue) population.

\begin{figure}[t!]
\resizebox{\hsize}{!}{\includegraphics{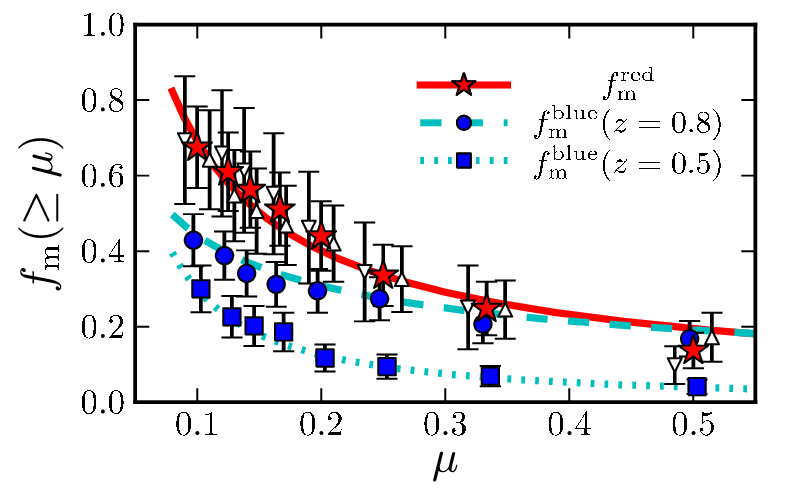}}
\caption{Merger fraction versus luminosity ratio in $B-$band, $\mu$. Stars, triangles and inverted triangles are the merger fraction of red primaries, $f_{\rm m}^{\rm red}$, at $z \in [0.2,0.95)$, $z = 0.8$, and $z = 0.5$, respectively. Dots and squares are the merger fraction of blue primaries, $f_{\rm m}^{\rm blue}$, at $z = 0.8$ and $z = 0.5$, respectively. The points are shifted when necessary to avoid overlap. The lines are the GLS fits of a power-law function, $f_{\rm m}\,(\geq\,\mu) \propto \mu^{s}$, to the combined $f_{\rm m}^{\rm red}$ ($s = -0.79$; solid), $f_{\rm m}^{\rm blue}$ at $z = 0.8$ ($s = -0.52$; dashed), and $f_{\rm m}^{\rm blue}$ at $z = 0.5$ data ($s = -1.26$; dotted). [{\it A colour version of this plot is available at the electronic edition}].}
\label{ffcolfig}
\end{figure}

With the previous definitions, the  principal sample comprises 268 red and 743 blue sources. We look for $r_{\rm p}^{\rm max} = 100h^{-1}$ kpc close companions, regardless of their colour, to ensure good statistics. As was mentioned in the previous section, the trends obtained with this search radius are representative to trends observed at smaller separation. We find that:
\begin{itemize}
\item The merger fraction of red galaxies ($f_{\rm m}^{\rm red}$; Table~\ref{ffmuredtab}) is higher than the merger fraction of blue galaxies ($f_{\rm m}^{\rm blue}$; Table~\ref{ffmubluetab}). For major mergers at $z = 0.8$, both fractions are comparable.
\item $f_{\rm m}^{\rm red}$ evolves little, if any, with cosmic time. Because of this lack of evolution, and to obtain better statistics, we combine both redshift ranges in the following (fourth column in Table~\ref{ffmuredtab}, and Fig.~\ref{ffcolfig}). We find that the power-law index is $s = -0.79 \pm 0.12$ in the range $0.2 \leq z < 0.95$. This implies that red galaxies have a similar number of minor and major companions, $f_{m/M}^{\rm red} = 1.06\pm0.22$.
\item $f_{\rm m}^{\rm blue}$ is lower at $z = 0.5$ than at $z = 0.8$. The observed evolution is faster for higher values 
of $\mu$ (Fig.~\ref{ffcolfig}), so we obtain different ($>2\sigma$) values for the power-law index: $s = -0.52 \pm 0.10$ 
at $z = 0.8$ and $s = -1.26 \pm 0.20$ at $z = 0.5$. The ratio of minor-to-major companions of blue galaxies grows from $f_{m/M}^{\rm blue} = 0.61\pm0.15$ at $z = 0.8$ to $f_{m/M}^{\rm blue} = 2.17\pm0.57$ at $z = 0.5$.
\end{itemize}

The fraction of principal galaxies that have a companion and are blue, $f_{\rm blue,1} = N_{\rm p}^{\rm blue}/N_{\rm p}$, does not depend on $\mu$ at $z = 0.8$, $f_{\rm blue,1} \sim 70$\%. On the other hand, $f_{\rm blue,1}$ increases when $\mu$ decreases at $z = 0.5$, varying from $f_{\rm blue,1} \sim 50$\% at $\mu \geq 1/10$ to $f_{\rm blue,1} \sim 40$\% at $\mu \geq 1/4$, in contrast with $\sim 70$\% at $z = 0.8$. The fraction of principal galaxies that have a companion and are red is $f_{\rm red,1} = N_{\rm p}^{\rm red}/N_{\rm p} = 1 - f_{\rm blue,1}$.

\begin{table}
\caption{Merger fraction of $L_{B,1} \gtrsim L_{B}^{*}$, red ($M_{NUV} - M_{r} \geq 4.25$) galaxies as a function of luminosity ratio $\mu$ for $r_{\rm p}^{\rm max} = 100h^{-1}$ kpc}
\label{ffmuredtab}
\begin{center}
\begin{tabular}{lcccc}
\hline\hline\noalign{\smallskip}
$\mu$ & $z = 0.5$ & $z = 0.8$ & $N_{\rm p}^{\rm red}$ & $z \in (0.2,0.95]$\\
\noalign{\smallskip}
\hline
\noalign{\smallskip}
1/2  & $0.098 \pm 0.050$ & $0.174 \pm 0.065$ &  8 & $0.137 \pm 0.047$\\
1/3  & $0.251 \pm 0.111$ & $0.246 \pm 0.077$ & 14 & $0.248 \pm 0.071$\\
1/4  & $0.345 \pm 0.131$ & $0.324 \pm 0.087$ & 19 & $0.336 \pm 0.081$\\
1/5  & $0.462 \pm 0.148$ & $0.418 \pm 0.101$ & 25 & $0.440 \pm 0.092$\\
1/6  & $0.552 \pm 0.160$ & $0.467 \pm 0.106$ & 29 & $0.511 \pm 0.097$\\
1/7  & $0.614 \pm 0.165$ & $0.523 \pm 0.113$ & 32 & $0.563 \pm 0.101$\\
1/8  & $0.659 \pm 0.167$ & $0.562 \pm 0.121$ & 35 & $0.610 \pm 0.104$\\
1/10 & $0.694 \pm 0.169$ & $0.655 \pm 0.131$ & 39 & $0.675 \pm 0.108$\\
\hline
\end{tabular}
\end{center}
\end{table}

\begin{table}
\caption{Merger fraction of $L_{B,1} \gtrsim L_{B}^{*}$, blue ($M_{NUV} - M_{r} < 4.25$) galaxies as a function of luminosity ratio $\mu$ for $r_{\rm p}^{\rm max} = 100h^{-1}$ kpc}
\label{ffmubluetab}
\begin{center}
\begin{tabular}{lccccc}
\hline\hline\noalign{\smallskip}
$\mu$ & \multicolumn{2}{c}{$z = 0.5$} & & \multicolumn{2}{c}{$z = 0.8$}\\
\noalign{\smallskip}
\cline{2-3} \cline{5-6}
\noalign{\smallskip}
      & $N_{\rm p}^{\rm blue}$ & $f_{\rm m}^{\rm blue}$ & & $N_{\rm p}^{\rm blue}$ & $f_{\rm m}^{\rm blue}$\\
\noalign{\smallskip}
\hline
\noalign{\smallskip}
1/2  &  4 & $0.041 \pm 0.019$ & & 16 & $0.168 \pm 0.047$ \\
1/3  &  6 & $0.069 \pm 0.027$ & & 20 & $0.207 \pm 0.051$ \\
1/4  &  8 & $0.094 \pm 0.032$ & & 27 & $0.274 \pm 0.057$ \\
1/5  & 10 & $0.117 \pm 0.036$ & & 30 & $0.295 \pm 0.058$ \\
1/6  & 14 & $0.186 \pm 0.051$ & & 32 & $0.312 \pm 0.059$ \\
1/7  & 15 & $0.202 \pm 0.053$ & & 35 & $0.341 \pm 0.061$ \\
1/8  & 17 & $0.226 \pm 0.055$ & & 39 & $0.388 \pm 0.064$ \\
1/10 & 23 & $0.300 \pm 0.062$ & & 41 & $0.429 \pm 0.069$ \\
\hline
\end{tabular}
\end{center}
\end{table}

We find that the fraction of companions that 
are blue is $f_{\rm blue,2} \sim 0.8$, regardless either of the colour of the principal or $\mu$. This means that 
red--red (dry), red--blue or blue--red (mixed), and blue--blue (wet) pairs account for $\sim$10\%/40\%/50\% of the pairs with a
minor companion in all the redshift range under study. This lack of evolution contrasts with the strong evolution 
of major mergers, for which the relative fractions are $\sim$5\%/40\%/55\% at $z = 0.8$ (similar to the minor ones), 
and $\sim$10\%/60\%/30\% at $z = 0.5$. From $z\sim0.8$ to $z\sim0.5$, the fraction of wet major mergers decreases 
by a factor of two, while dry and mixed mergers 
increase their importance. Our major merger trends are in agreement with \citet{deravel09} using an expanded data set, as well as previous works, e.g.,
\citet{lin08,bundy09}. These results show that the relative fraction of dry and mixed major mergers become more important with 
cosmic time for $L_{B} \gtrsim L_{B}^{*}$ galaxies in our redshift range due to the lack of blue primaries with major companions at low redshift, rather than from an increase in the major merger fractions of red galaxies as also pointed out by \citet{lin08}.

Previous work finds that the major merger fraction from close pairs depends on mass, with more massive galaxies having higher merger fractions \citep{deravel09,bundy09}. If blue principal galaxies at $z = 0.8$ were more massive by a factor of 3 than at $z = 0.5$ because of our $B-$band luminosity selection, this would explain the observed trend in $f_{\rm m}^{\rm blue}$. Using stellar masses determined in \citet{pozzetti07}, we do not find a significant change (less than 0.1 dex) in the median mass of red, $\log\,(\overline{M_{\star, {\rm red}}}/M_{\odot}) \sim 10.8$, and blue, $\log\,(\overline{M_{\star, {\rm blue}}}/M_{\odot}) \sim 10.3$, principal galaxies. This supports that the observed trends reflect a real evolution in the merger properties of 
blue galaxies. In addition, our results imply that more massive (red) galaxies have higher merger fractions than 
lower mass (blue) galaxies, in agreement with \cite{deravel09} and \cite{bundy09}. The study of the major and minor merger fraction in mass selected galaxies is beyond the scope of the present paper, and we will address this issue in a future work.

\section{The minor merger rate of $L_{B} \gtrsim L_{B}^{*}$ galaxies}\label{mrpair}

\subsection{The minor merger rate of the full population}\label{mrpairfull}
Our goal in this section is to estimate the minor merger ($1/10 \leq \mu < 1/4$) rate of bright galaxies in 
the range $0.2 \leq z < 0.95$. In the following we name the {\it merger rate} the number of mergers per Gyr per galaxy, noted $R$. 
Because the parameters involved in the translation of the merger fraction to the merger rate are better constrained 
for major mergers, we estimate them first and then expand to the minor merger rate.

Following \citet{deravel09}, we define the major merger rate as
\begin{equation}
R_{\rm MM} = f_{\rm MM}\,C_{\rm p}\, C_{\rm m}\, T_{\rm MM}^{-1},\label{mrpar}
\end{equation}
where the factor $C_{\rm p}$ takes into account the lost companions in the inner $5h^{-1}$ kpc \citep{bell06} and 
the factor $C_{\rm m}$ is the fraction of the observed close pairs that finally merge in a typical timescale $T_{\rm MM}$. 
We take $C_{\rm p} = r_{\rm p}^{\rm max}/(r_{\rm p}^{\rm max}-5h^{-1}\ {\rm kpc})$. The typical merger timescale depends 
on $r_{\rm p}^{\rm max}$ and can be estimated by cosmological and $N$-body simulations. We compute 
the major merger timescales from the cosmological simulations of \citet{kit08}, based on the Millennium simulation 
\citep{springel05}. These major merger timescales, denoted $T_{\rm MM}^{K08}$, refer to major mergers ($\mu > 1/4$ 
in stellar mass), and depend mainly on $r_{\rm p}^{\rm max}$ and on the stellar mass of the principal galaxy, 
with a weak dependence on redshift in our range of interest (see \citealt{deravel09}, for details). 
Taking $\log\, (M_{\star}/M_{\odot}) = 10.7$ as the average stellar mass of our principal galaxies with a close companion, 
we obtain the values in Table~\ref{tmm} for $r_{\rm p}^{\rm max} = 30$, 50 and 100 $h^{-1}$ kpc, and $\Delta v^{\rm max} = 500$ km s$^{-1}$. 
In every case we assume an uncertainty of 0.2 dex in the mass of the principal galaxies to estimate the error 
in $T_{\rm MM}^{K08}$. These timescales already include the factor $C_{\rm m}$ \citep[see][]{patton08,bundy09,lin10}, 
so we take $C_{\rm m} = 1$ in the following. These timescales are for central - satellite mergers, and satellite - satellite pairs could have different timescales. However, only 1 of the 103 close pairs under study is satellite - satellite, so the use of principal - satellite timescales is justified. We also remark that the velocity condition $\Delta v^{\rm max} = 500$ km s$^{-1}$ selects close bound systems even when they are located in dense environments, but in these environments the probability of finding unbound close pairs increases. This is taken into account in the cosmological averaged merger timescales (see also \citealt{lin10}).

Since the assumed merger timescale is the most uncertain quantity in Eq.~(\ref{mrpar}), we compare $T_{\rm MM}^{K08}$ 
with other recent estimations in the literature. \citet{lotz10t} perform $N$-body/hydrodynamical simulations of major 
and minor mergers to study the merger timescales of morphological and close pair approaches. The principal galaxy in 
their simulations has $\log\, (M_{\star}/M_{\odot}) = 10.7$, similar to the average mass of our principal galaxies with a close companion, so 
their major merger timescales, denoted $T_{\rm MM}^{JL10}$, should be comparable to the previous $T_{\rm MM}^{K08}$. 
We summarize the average values of $T_{\rm MM}^{JL10}$ in Table~\ref{tmm} after correcting with the factor $C_{\rm p}$. 
We find that $T_{\rm MM}^{JL10} < T_{\rm MM}^{K08}$. However, the $T_{\rm MM}^{K08}$ include the factor $C_{\rm m}$, 
while the $T_{\rm MM}^{JL10}$ do not. Applying to $T_{\rm MM}^{JL10}$ a typical value of $C_{\rm m} = 0.6$ \citep{patton00,lin04,lin10,bell06}, we find that {\it both timescales agree and therefore yield similar merger rates}. On the other hand, \citet{lin10} use cosmological simulations to study $C_{\rm m}$ and the merger timescale, denoted $T_{\rm MM}^{LL10}$. They find $T_{\rm MM}^{LL10} \sim 1.4$ Gyr for $\log\, (M_{\star}/M_{\odot}) \sim 10.3$ galaxies and $r_{\rm p} \leq 50h^{-1}$ kpc (this value includes the factor $C_{\rm m} = 0.7$ derived from their simulations). This timescale is lower by a factor of two than the one from \citet{kit08} for this mass, $T_{\rm MM}^{K08} = 2.7$ Gyr. However, \citet{kit08} assume that the galaxy merger occurs a dynamical friction time after the dark matter halo merger; while \citet{lin10} do not consider this extra time. This fact mitigates the difference between both works, but a more detailed comparison is needed. In the following we omit the super index in $T_{\rm MM}^{\rm K08}$ for clarity.

The merger rate is an absolute quantity, and should not depend on the $r_{\rm p}^{\rm max}$ that we use to infer it. Because of this, the increase of the merger fraction with $r_{\rm p}^{\rm max}$ (Sect.~\ref{ffmu}, Fig.~\ref{brpfig}) must be compensated with the increase in $T_{\rm MM}$. For two different search radius, $r_{\rm p,1}^{\rm max}$ and $r_{\rm p,2}^{\rm max}$, this implies that
\begin{equation}
\Delta T_{\rm MM}\,(r_{\rm p,1}^{\rm max},r_{\rm p,2}^{\rm max}) = \frac{T_{\rm MM}\,(r_{\rm p,1}^{\rm max})}{T_{\rm MM}\,(r_{\rm p,2}^{\rm max})} = \frac{C_{\rm p,1}}{C_{\rm p,2}} \bigg(\frac{r_{\rm p,1}^{\rm max}}{r_{\rm p,2}^{\rm max}}\bigg)^{q}.
\end{equation}
From our observational results we infer that $\Delta T_{\rm MM} (50,30) = 1.5$ and $\Delta T_{\rm MM} (100,50) = 1.8$. These values compare nicely with the ratios from Table~\ref{tmm} timescales, $\Delta T_{\rm MM} (50,30) = 1.6$ and $\Delta T_{\rm MM} (100,50) = 1.8$. This supports the robustness of the assumed $T_{\rm MM}$, although the normalization of these timescales have a factor of two uncertainty. We estimate the final major merger rate averaging the values derived from the 30, 50 and $100h^{-1}$ kpc merger fractions, and its error as the average of the individual merger rates' errors.

\begin{table}

\caption{Major merger timescales of $L_{B,1} \gtrsim L^{*}_{B}$ galaxies}
\label{tmm}
\begin{center}
\begin{tabular}{lccc}
\hline\hline\noalign{\smallskip}
$r_{\rm p}^{\rm max}$ & $T_{\rm MM}^{K08}$ & $T_{\rm MM}^{JL10}$ &  $T_{\rm MM}^{JL10}/C_{\rm m}$\\
($h^{-1}$ kpc) & (Gyr) & (Gyr) &  (Gyr)\\
\noalign{\smallskip}
\hline
\noalign{\smallskip}
30  & $1.4 \pm 0.2$ & 0.9 & 1.5\\
50  & $2.3 \pm 0.3$ & 1.5 & 2.5\\
100 & $4.2 \pm 0.5$ & 2.4 & 4.0\\
\hline
\end{tabular}
\end{center}
\end{table}
 
We obtain the minor merger rate, defined as the merger rate of $1/10 \leq \mu < 1/4$ close pairs, from the major one as
\begin{equation}
R_{\rm mm} = f_{m/M} \frac{R_{\rm MM}}{\Upsilon}\label{mrmm},
\end{equation}
where the factor $\Upsilon$ accounts for the difference in the minor merger timescale with respect to the major merger one
in close pairs, $T_{\rm mm} = \Upsilon \times T_{\rm MM}$. Only a few studies in the literature attempt to estimate 
$\Upsilon$: \citet{jiang08} study the merger timescale of dark matter haloes, finding $\Upsilon \sim 2$. On the other 
hand, \citet{lotz10t} obtain $\Upsilon = 1.5 \pm 0.1$ from $N$-body/hydrodynamical simulations. As we have already shown, 
the major merger timescales from \citet{lotz10t} are similar to ours, so we assume the minor-to-major merger time 
scale from \citet{lotz10t} in the following. We also assume that the factor $C_{\rm m}$ for minor mergers is the same 
as the one for major mergers.

\begin{table*}
\caption{Minor, major and total merger rate of $L_{B} \gtrsim L^{*}_{B}$ galaxies}
\label{mrtab}
\begin{center}
\begin{tabular}{lcccccc}
\hline\hline\noalign{\smallskip}
Merger rate & \multicolumn{2}{c}{All galaxies} & Red galaxies & \multicolumn{2}{c}{Blue galaxies}\\
\noalign{\smallskip}
\cline{2-3} \cline{5-6}
\noalign{\smallskip}
$($Gyr$^{-1})$ & $z = 0.50$ & $z = 0.80$ & $z \in [0.2,0.95)$ & $z = 0.50$ & $z = 0.80$  \\
\noalign{\smallskip}
\hline
\noalign{\smallskip}
$R_{\rm MM}$ & $ 0.044 \pm 0.016$ & $ 0.070 \pm 0.021$ & $0.091 \pm 0.025$ & $ 0.021 \pm 0.007$ & $ 0.060 \pm 0.014$\\
$R_{\rm mm}$ & $ 0.045 \pm 0.019$ & $ 0.034 \pm 0.012$ & $0.064 \pm 0.022$ & $ 0.030 \pm 0.013$ & $ 0.024 \pm 0.008$ \\
$R_{\rm m}$  & $ 0.089 \pm 0.025$ & $ 0.104 \pm 0.025$ & $0.155 \pm 0.033$ & $ 0.051 \pm 0.015$ & $ 0.084 \pm 0.016$\\
\hline
\end{tabular}
\end{center}
\end{table*}

\begin{figure}[t!]
\resizebox{\hsize}{!}{\includegraphics{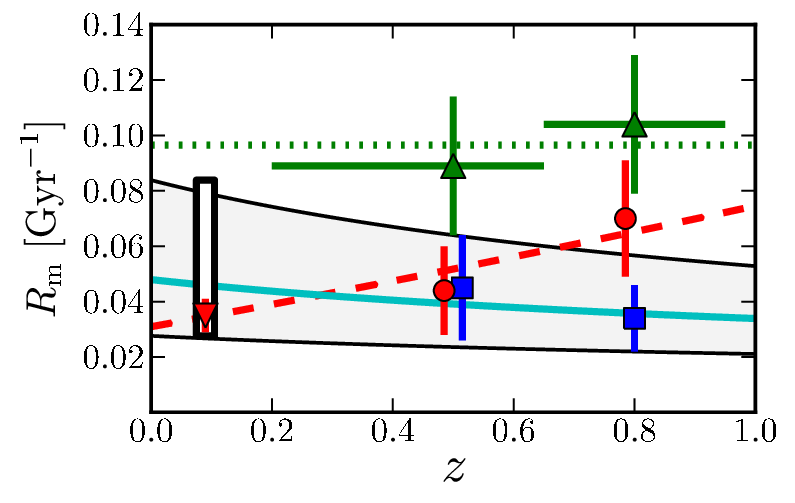}}
\caption{Merger rate of $M_{B}^{\rm e} \leq -20$ galaxies versus redshift. Dots are the major merger rate ($\mu \geq 1/4$), squares are the minor merger rate ($1/10 \leq \mu < 1/4$), and  triangles are the total (major + minor, $\mu \geq 1/10$) merger rate. The points are shifted when necessary to avoid overlap. The $z$ error bars in the total merger rate mark the redshift range spanned by VVDS-Deep data. The inverted triangle is the major merger rate of $M_B^{\rm e} \leq -20$ galaxies from MGC at $z = 0.09$. The white rectangle identifies the local ($z = 0.09$) minor merger fraction estimated from the total and the major merger ones, while the gray area marks the most probable minor merger rate values in the range $0 < z < 1$ (see text for details). The solid line is the best fit of a power-law function with a fixed index, $f_{\rm mm} \propto (1+z)^{-0.5}$, to the minor merger rate data. The dashed line 
is the least-squares fit of a power-law function to the major merger rate data. The dotted line is the major + minor merger rate if it is assumed constant. [{\it A colour version of this plot is available at the electronic edition}].}
\label{mrfig}
\end{figure}

Finally, the total merger rate is $R_{\rm m} = R_{\rm MM} + R_{\rm mm}$. We summarize our results on the merger rates in 
Table~\ref{mrtab}, and we show them in the Fig.~\ref{mrfig}. We find that
\begin{enumerate}

\item {\it The minor merger rate $R_{\rm mm}$ ($1/10 \leq \mu < 1/4$) decreases with increasing redshift}, although our measurements are consistent with a constant minor merger rate within errors. 
We further discuss the evolution of $R_{\rm mm}$  in Sect.~\ref{mrmmz}. This is the first quantitative measurement of the minor merger rate using close pair statistics at these redshifts.

\item {This trend is clearly different from the evolution of the major merger rate} ($\mu \geq 1/4$) which we find
is increasing with redshift, in agreement with \citet{deravel09}, and to  previous studies in the literature 
\citep[e.g.,][]{lefevre00,conselice03ff,conselice09cos,clsj09ffgoods,bridge10}. 

\item The total merger rate (major + minor) is consistent either with a mild increase with redshift or with a constant $R_{\rm m} \sim 0.1$ Gyr$^{-1}$. 
\end{enumerate}

\subsection{The minor merger rate of red and blue galaxies}
We apply the steps in the previous section to estimate the major, minor and total merger rate of red and blue 
galaxies. We take $T_{\rm MM}^{\rm red} = 3.9$ Gyr and $T_{\rm MM}^{\rm blue} = 4.8$ Gyr for $r_{\rm p}^{\rm max} = 100h^{-1}$ kpc because of the different 
average stellar mass of red and blue principal galaxies, while the factor $\Upsilon$ does not depend on the gas content of 
the galaxies \citep{lotz10gas}. The merger rates that we obtain are listed in Table~\ref{mrtab}. The merger rates (minor and major) 
of red galaxies do not evolve with redshift in the range under study, $R_{\rm mm}^{\rm red} = 0.064$ Gyr$^{-1}$ 
and $R_{\rm MM}^{\rm red} = 0.091$ Gyr$^{-1}$. \citet{gongar09} find that the minor and major merger rate of Elliptical 
Like Objects (ELOs) at $z \sim 0.75$ in their cosmological simulations are $R_{\rm mm} = 0.06$ Gyr$^{-1}$ and 
$R_{\rm MM} = 0.08$ Gyr$^{-1}$, in good agreement with our observed values. On the other hand, \citet{stewart09} 
model predicts that $R_{\rm mm} \sim R_{\rm MM}$ for $\mu_{\rm MM} = 1/3$ (see also \citealt{hopkins09bulges}), 
while from our observations we infer $R_{\rm mm} = 1.1 \times R_{\rm MM}$ for $\mu_{\rm MM} = 1/3$. 

The minor merger rate of blue galaxies, denoted $R_{\rm mm}^{\rm blue}$, increases by $\sim$20\% from $z = 0.8$ to 
$z = 0.5$, but the measured values are compatible with a constant merger rate within error bars, $R_{\rm mm}^{\rm blue} \sim 0.027$ Gyr$^{-1}$. To the contrary, 
the major merger rate, denoted $R_{\rm MM}^{\rm blue}$, decreases by a factor of three from $z = 0.8$ to $z = 0.5$,
as noted by \citet{deravel09}. These trends suggest that the stability or increase with cosmic time of the minor merger rate 
found in the previous section is a consequence of the 
evolution in the fraction of bright galaxies that are red: as time goes by, the red fraction increases 
\citep[e.g.,][]{fontana09, ilbert10}. Because the minor merger rate of red galaxies is a factor 
of $\sim2.5$ higher than the one of blue galaxies, and both are roughly constant, the increase in the red fraction 
implies an increase in the global (red+blue) minor merger rate. This effect is also present in the major merger rate, 
but in this case $R_{\rm MM}^{\rm blue}$ decreases with cosmic time, and the increase in the red fraction is only 
a mild evolution, as found by \citet{deravel09}.

\subsection{The volumetric minor merger rate}
The volumetric merger rate (i.e., the number of mergers per unit volume and time) is a complementary measure 
to the merger rate estimated in the previous sections. To obtain the volumetric merger rate, denoted 
$\Re$, we multiply the merger rate by the number density of all/red/blue galaxies with 
$M_B^{\rm e} \leq -20$ in VVDS-Deep at each redshift \citep{ilbert05}. We summarize the values 
of $\Re$ in Table~\ref{mrvtab}. All trends are similar to those found in the previous section. 
Interestingly, we find that 
$\Re_{\rm mm}^{\rm red} \sim \Re_{\rm mm}^{\rm blue} \sim 3.5 \times 10^{-5}\ {\rm Mpc}^{-3}\ {\rm Gyr}^{-1}$. 
The merger rate of red galaxies is higher by a factor of $\sim$ 2.5 than that of the blue ones, but the number density of the 
latter is higher than of the former, hence making the volumetric merger rates comparable.

\begin{table*}
\caption{Minor, major and total volumetric merger rate of $L_{B} \gtrsim L^{*}_{B}$ galaxies}
\label{mrvtab}
\begin{center}
\begin{tabular}{lcccccc}
\hline\hline\noalign{\smallskip}
Merger rate & \multicolumn{2}{c}{All galaxies} & Red galaxies & \multicolumn{2}{c}{Blue galaxies}\\
\noalign{\smallskip}
\cline{2-3} \cline{5-6}
\noalign{\smallskip}

$(\times 10^{-5}\ {\rm Mpc}^{-3}\ {\rm Gyr}^{-1})$ & $z = 0.50$ & $z = 0.80$ & $z \in [0.2,0.95)$ & $z = 0.50$ & $z = 0.80$  \\
\noalign{\smallskip}
\hline
\noalign{\smallskip}
$\Re_{\rm MM}$ & $  8.3 \pm 3.0$ & $ 12.6 \pm 3.8$ & $4.4 \pm 1.2$ & $ 2.8 \pm 1.0$ & $  8.3 \pm 1.9$\\
$\Re_{\rm mm}$ & $  8.6 \pm 3.6$ & $  6.1 \pm 2.2$ & $3.1 \pm 1.1$ & $ 4.0 \pm 1.8$ & $  3.4 \pm 1.2$\\
$\Re_{\rm m}$  & $ 16.8 \pm 4.7$ & $ 18.8 \pm 4.4$ & $7.5 \pm 1.6$ & $ 6.8 \pm 2.0$ & $ 11.7 \pm 2.3$\\
\hline
\end{tabular}
\end{center}
\end{table*}

\section{Discussion}\label{discussion}
In this section we estimate the evolution of the minor merger fraction and rate with redshift, and discuss the contribution of minor mergers to the evolution of bright galaxies since $z \sim 1$, comparing it to the contribution of major mergers.

\subsection{The evolution of the minor merger fraction with redshift}\label{ffmmevol}
The evolution of the merger faction with redshift up to $z \sim 1.5$ is well parametrized  by a power-law  \citep[e.g.,][]{lefevre00,clsj09ffgoods,deravel09},
\begin{equation}
f_{\rm m}(z) = f_{\rm m,0}\,(1+z)^{m}.\label{fmz}
\end{equation}
Our results alone suggest that the merger fraction evolves faster for higher $\mu$, with $m=5.6$ for equal 
luminosity companions ($\mu = 1$), $m = 2.4$ for major companions with $\mu \geq 1/4$, and $m = 0.8$ for major + minor 
companions ($\mu \geq 1/10$). This mild evolution in the total (major + minor) merger fraction is also suggested by the morphological studies of \citet{lotz08ff} and \citet{jogee09}.

To better constrain the evolution with redshift of the minor merger fraction, a local reference is important. \citet{darg10i} estimate that the minor merger fraction is similar to the major one ($f_{m/M} \sim 1$, $\mu \gtrsim 1/3$) in Galaxy Zoo\footnote{http://www.galaxyzoo.org} \citep{lintott08}; the latter is based on the visual classification of Sloan Digital Sky Survey (SDSS\footnote{http://sdss.org/}, \citealt{adelman06}) galaxies by internet users. However, their sample is incomplete for minor companions, so their $f_{m/M}$ is a lower limit. On the other hand, \citet{woods07} study the different properties of major ($\Delta m_{z} < 2$, $\mu \gtrsim 1/7$) and minor ($\Delta m_{z} > 2$, $\mu \lesssim 1/7$) close pairs in SDSS. Unfortunately, they do not attempt to derive merger fractions, but the influence of close companions on galaxy properties (see also \citealt{ellison08,patton11}). Therefore, to our knowledge, there does not seem to be any local estimation of the minor merger fraction of bright galaxies in the literature. As a close proxy, we estimate the local merger fraction as $f_{\rm mm} = f_{\rm m}\,(\mu\,\geq 1/10) - f_{\rm MM}$. We follow the methodology in Sect.~\ref{ncs} to measure the major ($\mu\,\geq 1/4$) merger fraction of $M_{B}^{\rm e} \leq -20$ galaxies at $z = 0.09$ from the Millennium Galaxy Catalogue (MGC\footnote{http://eso.org/$\sim$jliske/mgc/}, \citealt{liske03}). This survey comprises 10095 galaxies with $B_{MGC} < 20$ over 37.5~$\deg^2$, with a spectroscopic completeness of 96\% (\citealt{driver05}; see also \citealt{depropris05,depropris07}). We obtain $f_{\rm MM}^{\rm MGC} = 0.139\pm0.009$ for $r_{\rm p}^{\rm max} = 100h^{-1}$ kpc. We then assume two different types of evolution for the major + minor merger fraction: (1) a constant evolution with redshift, $f_{\rm m}\,(\mu\,\geq 1/10) = 0.461$ for $r_{\rm p}^{\rm max} = 100h^{-1}$ kpc, which implies $f_{\rm mm}(0.09) = 0.322$; and (2) an evolution which evolves with redshift as $m = 0.8$ (fit of a power-law function to our observational major + minor merger fractions), which implies $f_{\rm mm}(0.09) = 0.187$. Finally, we fit Eq.~(\ref{fmz}) to our minor merger fraction data and both local estimates, defining a confidence area for the minor merger fraction between $z = 0$ and $z = 1$ (Fig.~\ref{ffmmfig}). This area is limited by the following curves,
\begin{eqnarray}
f_{\rm mm}^{\rm up} =  0.393\,(1+z)^{-0.32},\\
f_{\rm mm}^{\rm down} =  0.182\,(1+z)^{-0.25}.
\end{eqnarray}
The power law-index from the fits is $m = -0.4 \pm 0.7$. The negative value implies that the minor merger fraction decreases with increasing redshift. We note that our results are compatible with a constant $f_{\rm mm}$ since $z = 1$ (i.e., $m = 0$). Even in that case, the minor merger fraction does not evolve in the same way as the major one, that increases 
with redshift ($m > 0$, see below). \citet{abbas10} use Halo Occupation Distribution (HOD) models to interpret the evolution since $z \sim 1$ of the correlation function from VVDS-Deep \citep[see also][]{lefevre05cluster} and SDSS. Their results suggest that the average number of satellite galaxies per dark matter halo increases with cosmic time, which could be related with our suggested increase in the minor merger fraction. Specifically, we expect the minor merger fraction in the local universe to be two to three times the major merger one. Direct measurements of the minor merger fraction at low redshift will be needed to better constrain the minor merger fraction evolution with $z$.

The least-squares fit to the major merger data yields (Fig.~\ref{ffmmfig})
\begin{equation}
f_{\rm MM} =  (0.116 \pm 0.024)\,(1+z)^{1.3 \pm 0.5}.
\end{equation}
In a previous work in VVDS-Deep, \citet{deravel09} measured the major merger fraction ($\mu \geq 1/4$) of less luminous galaxies than those reported in present paper. They find that the major merger fraction evolves faster with $z$ for fainter samples, with a power-law index $m = 4.7$ for $M_B^{\rm e} \leq-18$ galaxies and $m = 3.1$ for $M_B^{\rm e} \leq-18.77$ galaxies. The evolution of $m = 1.3$ for the major merger fraction of $M_B^{\rm e} \leq-20$ galaxies confirms the trend found by \citet{deravel09} and extends it to brighter galaxies.

\subsection{The evolution of the power-law index $s$ with redshift}
In a previous study, \citet{clsj10pargoods} have attempted to measure the power-law index $s$. They 
find $s \sim -0.6$ at $z \in [0.2,1.1)$ for principal galaxies with $M_{\star} \gtrsim 10^{10}\ M_{\sun}$. 
This value is similar to ours at $z = 0.8$, but at $z \sim 0.5$ the discrepancy between both studies is 
important ($>2\sigma$). This suggests that $s$ depends not only on both redshift and colour, but also on 
stellar mass. Because the $B$-band luminosities of red galaxies are only slightly affected by star formation, 
our red merger fraction is a proxy of the merger fraction of $\log\, (M_{\star}/M_{\odot}) \sim 10.8$ galaxies. 
We therefore find that the power-law index does not evolve for massive galaxies, $s = -0.79\pm0.12$. This, 
combining with \citet{clsj10pargoods} results, suggests that (i) $s$ does not evolve with $z$ in mass-selected 
samples; that is, the evolution of the total (major + minor) merger fraction is similar to that of the major merger one, as predicted by the cosmological models of \citet{stewart09}, and (ii) the power-law index is lower for massive galaxies indicating that massive galaxies 
have a higher minor-to-major merger ratio than less massive ones. The minor merger fraction in different 
mass-selected samples will be the subject of a future work to expand on results presented here.

\subsection{The redshift evolution of the minor merger rate}\label{mrmmz}
Similarly to the minor merger fraction, there does not seem to exist any published reference in
the refereed literature for the local minor merger rate. 
We follow the same steps as in Sect.~\ref{ffmmevol} to estimate a confidence area for the minor merger rate in the range $0 < z < 1$. The major merger rate in the MCG at $z = 0.09$ is $R_{\rm MM}^{\rm MGC} = 0.035\pm0.006$ Gyr$^{-1}$, while the confidence area is limited by the following curves (Fig.~\ref{mrfig}),
\begin{eqnarray}
R_{\rm mm}^{\rm up} =  0.084\,(1+z)^{-0.67},\\
R_{\rm mm}^{\rm down} =  0.028\,(1+z)^{-0.39}.
\end{eqnarray}
The power law-index inferred from the fits is $n = -0.5 \pm 0.7$. As in Sect.~\ref{ffmmevol}, a negative power-law index for $R_{\rm mm}$ implies that the minor merger rate decreases with redshift. Also in this case the value of $n$ is compatible with a constant minor merger rate ($n = 0$), but again its evolution is different than that of the major merger rate, that increases 
with redshift ($n > 0$, see below). A local reference is needed to better constraint the evolution of $R_{\rm mm}$. If we repeat this study with the volumetric merger rate, the confidence area is limited by
\begin{eqnarray}
\Re_{\rm mm}^{\rm up} =  11.3\,(1+z)^{0.19}\ \times 10^{-5}\ {\rm Mpc}^{-3}\ {\rm Gyr}^{-1},\\
\Re_{\rm mm}^{\rm down} = 6.8\,(1+z)^{-0.91}\ \times 10^{-5}\ {\rm Mpc}^{-3}\ {\rm Gyr}^{-1}.
\end{eqnarray}
In this case the evolution is $n = -0.5 \pm 0.7$.

The fit to both major merger rates is
\begin{eqnarray}
R_{\rm MM} =  (0.031 \pm 0.006)\,(1+z)^{1.3 \pm 0.6},\\
\Re_{\rm MM} =  (6.6 \pm 1.2)\,(1+z)^{0.9 \pm 0.4}\ \times 10^{-5}\ {\rm Mpc}^{3}\ {\rm Gyr}^{-1}.
\end{eqnarray}
\citet{deravel09} estimate the volumetric major merger rate ($\mu \geq 1/4$) finding, as for the merger fraction, 
that it evolves faster for fainter samples, with a power-law index $n = 2.2$ for $M_B^{\rm e} \leq-18$ galaxies and $n = 1.6$ for $M_B^{\rm e} \leq-18.77$ galaxies, so our $n = 0.9$ follows the trend of decreasing $n$ for brighter galaxies found by \citet{deravel09}. On the other hand, the volumetric merger rate of $M_B^{\rm e} \leq-18$ galaxies is a factor of $\sim5$ higher than the one of $M_B^{\rm e} \leq-20$ galaxies. This is because the number density is lower for bright galaxies than for the fainter ones. The same trend is observed in mass-selected samples \citep{clsj09ffgs}.

\subsection{The role of minor mergers in the mass assembly of luminous galaxies}
We can obtain the average number of mergers per galaxy between $z_2$ and $z_1 < z_2$ as
\begin{equation}
N_{\rm m} = \int_{z_1}^{z_2} R_{\rm m}\frac{\rm{d}z}{(1+z)H_0E(z)},\label{numm}
\end{equation}
where $E(z) = \sqrt{\Omega_{\Lambda} + \Omega_{m}(1+z)^3}$ in a flat universe. 
The definitions of $N_{\rm MM}$ and $N_{\rm mm}$ are analogous. Using results from the previous section, 
we obtain $N_{\rm m} = 0.73\pm0.21$, with $N_{\rm MM} = 0.37\pm0.13$ and 
$N_{\rm mm} = 0.36\pm0.17$ from $z=1$ to $z=0$, indicating that {\it the number of 
minor mergers per bright galaxy since $z = 1$ is similar to the number of major ones}. Note that these values and those reported in the following have an additional factor of two uncertainty due to the merger timescales derived from simulations (Sect.~\ref{mrpairfull}). In their work, \citet{pozzetti10} find that almost all the evolution in the stellar mass function since $z \sim 1$ is consequence of the observed star formation \citep[see also][]{vergani08}, and estimate $N_{\rm m} \sim 0.7$ mergers since $z \sim 1$ per $\log\,(M_{\star}/M_{\odot}) \sim 10.6$ galaxy, similar to the average mass of our $M_B^{\rm e} \leq -20$ galaxies, are needed to explain the remaining evolution. Their result agrees with our direct estimation, but they infer $N_{\rm MM} < 0.2$. This value is half of ours, pointing out that close pair studies are needed to understand accurately the role of major/minor mergers in galaxy evolution.

In addition to the mean number of mergers per galaxy, we have estimated the mass accreted by bright galaxies 
since $z = 1$ due to major and minor mergers. For this, we take $\mu$ as a proxy of the mass ratio between the galaxies in the pair. We can determine the mean merger ratio of major ($\overline{\mu_{MM}}$), and minor mergers ($\overline{\mu_{mm}}$) as
\begin{eqnarray}
\overline{\mu_{MM}} = \frac{s}{s+1}\frac{1-\mu_{\rm MM}^{s+1}}{1-\mu_{\rm MM}^{s}},\label{meanMM}\\
\overline{\mu_{mm}} = \frac{s}{s+1}\frac{\mu_{\rm mm}^{s+1}-\mu_{\rm MM}^{s+1}}{\mu_{\rm mm}^{s}-\mu_{\rm MM}^{s}}.\label{meanmm}
\end{eqnarray}
For $\mu_{\rm MM} = 1/4$ and $\mu_{\rm mm} = 1/10$ we obtain $\overline{\mu_{\rm MM}} = 0.47$ and $\overline{\mu_{\rm mm}} = 0.15$, values that depend slightly on $s$: 
the mean merger ratios change less than $10$\% in the range probed by our results, $s \in [-1.25,-0.58]$. 
We assume these values of $\overline{\mu_{\rm MM}}$ and $\overline{\mu_{\rm mm}}$ hereafter. Weighting the number 
of mergers with its corresponding merger ratio, we infer that {\it mergers of companions with $\mu$
in the range $1/10$ to $1$ increase the mass of bright galaxies since $z = 1$ by $23\pm8$\%}. 
We further infer that the relative contribution of major and minor mergers to this mass 
assembly is 75\% and 25\%, respectively. Because the factor of two uncertainty in the merger timescales affects in the same way major and minor mergers, this relative contribution is a robust result. In their cosmological models, \citet{hopkins10fusbul} predict that the relative contribution of major and minor mergers in the spheroids assembly of $\log\,(M_{\star}/M_{\odot}) \sim 10.6$ galaxies is $\sim80\%$/20\%, in good agreement with our observational result.

Therefore, we have demonstrated that minor mergers do contribute to the mass assembly of bright galaxies, at a level corresponding to about a third of the major mergers contribution.

\subsection{Mergers and the evolution of red galaxies since $z \sim 1$}\label{redmerger}
Because the merger properties of red and blue galaxies are very different, we estimate here the role of minor and 
major mergers in the evolution of red galaxies since $z \sim 1$. We assume a constant major and minor merger rate for red galaxies from $z = 0$ to $z = 1$, as found in Section \ref{mrpair}. 
Applying Eq.~(\ref{numm}) to $R_{\rm mm}^{\rm red}$ and $R_{\rm MM}^{\rm red}$, we obtain 
that {\it the average number of mergers per red galaxy since $z = 1$ is $N_{\rm m}^{\rm red} = 1.2\pm0.3$}, 
with $N_{\rm MM}^{\rm red} = 0.7\pm0.2$ and $N_{\rm mm}^{\rm red} = 0.5\pm0.2$. These values are 
higher than those from the global population, reflecting the higher merger rate of red galaxies.

We find that red galaxies of $\log\, (M_{\star}/M_{\odot}) \sim 10.8$ have undergone $\sim 1.2$ merger 
events since $z \sim 1$, but it is important to quantify the impact of mergers in the mass assembly 
of these galaxies. Weighting the number of mergers with their corresponding mean merger ratio 
(Eqs.~[\ref{meanMM}] and [\ref{meanmm}]), we find that {\it mergers can increase $40\pm10$\%
the mass of red galaxies since $z = 1$}. Because blue companions have a lower mass-to-light ratio than the red ones, this mass increase is an upper limit. The relative contribution of major/minor mergers to this mass 
assembly is 80\%/20\%, indicating that the mass of red galaxies increases by $\sim 10$\% since $z = 1$ due to minor mergers. 

Several authors have studied the luminosity function (LF) and the clustering to 
constrain the evolution of luminous red galaxies (LRGs) with redshift. 
They find that the bright end ($L \gtrsim 2.5L^{*}$) of 
the LF is mostly in place since $z \sim 0.8$ \citep[e.g.,][]{zucca06,brown07,scarlata07ee}. Since LRGs have a 
negligible star formation \citep{roseboom06}, the evolution of the bright end of the LF, if any, must be 
due to mergers. \citet{brown08} find that bright LRGs ($M_B \lesssim -21.8 \sim 4L^{*}$) have increased their 
mass $\sim30$\% since $z = 1$ (see also \citealt{brown07}), in agreement with our result. \citet{cool08} state 
that $L > 3L^{*}$ galaxies have increased their stellar mass less than 50\% since $z \sim 0.9$, an upper limit 
also consistent with our measurement.  
On the other hand, \citet{vandokkum10} study the evolution of massive galaxies with $\log\, (M_{\star}/M_{\odot}) \gtrsim 11.3$ since $z \sim 2$, inferring that they increase their mass 
$\sim 40$\% since $z \sim 1$ to the present by mergers (i.e., their star-formation is negligible in that redshift range, see also \citealt{walcher08} and \citealt{drory08}), 
in good agreement with our direct measurement. Although the stellar mass and luminosity range probed by 
\citet{vandokkum10} and previous LF works is $\sim$3 times higher than ours, and we use $B-$band luminosity as 
a proxy of mass, the agreement with these studies is remarkable and supports that mergers are an important
contributor to the evolution of the most massive red galaxies since $z \sim 1$.

While mergers directly increase the mass in red galaxies, they also modify their inner structure. It is now well established 
that massive, $\log\, (M_{\star}/M_{\odot}) \gtrsim 11$, early-type galaxies have, on average, lower effective 
radius ($r_{\rm e}$) at high redshift than locally, being $\sim 2$ to $\sim 4$ times smaller at $z \sim 1$ and $z \sim 2$, respectively \citep{daddi05,trujillo06,trujillo07,buitrago08,vandokkum08,vandokkum10,vanderwel08esize,toft09,williams10}. 
These high-redshift compact galaxies are sparse in the local universe \citep{trujillo09,taylor10}, implying that they 
evolve since $z \sim 2$ to the present. It has been suggested that compact galaxies are the cores of present day 
ellipticals, and that they increase their size by adding stellar mass in the outskirts of the galaxy 
\citep{bezanson09,hopkins09core,vandokkum10}. Equal-mass mergers ($\mu = 1$) are efficiently increasing the mass 
of the galaxies, but not their size ($r_{\rm e} \propto M_{\star}$); while for un-equal mass mergers ($\mu < 1$) 
the size increase is higher for the same accreted mass ($r_{\rm e} \propto M_{\star}^{2}$; \citealt{bezanson09,hopkins10size}). 
We find that red galaxies increase their mass $\sim40$\% since $z \sim 1$ due primarily to un-equal mass mergers. 
This corresponds to a size increase by a factor of $\sim2$, which is similar to the growth derived by size studies. 
Our results therefore suggest that un-equal mass mergers ($\mu < 1$) could be the 
dominant process in the size growth of massive galaxies since $z \sim 1$, as predicted by the cosmological simulations 
of \citet{naab09} or \citet{hopkins10size}. Future studies of the merger fraction as a function of the size of galaxies 
are needed to better understand the evolution of compact galaxies.

\citet{kaviraj10} found that $\sim 30$\% of early types at $0.5 < z < 0.7$ present distorted morphologies. 
This fraction is $\sim25$\% if we restrict the analysis to $M_V \lesssim -21.5$ galaxies (this selects 
$M_{B}^{\rm e} \lesssim -20$ galaxies at $z = 0.6$ assuming $B-V = 0.7$, the main $M_B - M_V$ colour of our red 
galaxies in the range $0.5 < z < 0.7$). Interestingly, \cite{conselice07} also found that $\sim25$\% of the 
early-types with $\log\,(M_{\star}/M_{\odot}) \geq 10.8$ in the Palomar/DEEP2 survey present signs of interactions at these redshifts. 
If we assume a visibility timescale of $T_{\rm dET}\sim 1$ Gyr for \citet{kaviraj10} distorted early-types (dET), 
we need a total (major + minor) merger rate of $R_{\rm dET} \sim 0.25$ Gyr$^{-1}$ to explain the observed fraction of dET. 
This value is higher than our red merger rate, $R_{\rm m}^{\rm red} = 0.155 \pm 0.033$ Gyr$^{-1}$, but we infer and additional $R_{\rm dET}^{\rm blue} \sim 0.1$ from the major merger rate of blue galaxies, that can also lead to dET (Sect.~\ref{bluemerger}). Mergers could therefore be common enough to explain the observed frequency of dET at $z = 0.6$, 
with minor mergers accounting for $\sim30$\% of the observed dET. $N$--body simulations are needed to better 
determine $T_{\rm dET}$ and the minimum $\mu$ that produces observable tidal features. We also note that minor mergers
with luminosity or mass ratios less than 1/10 may also contribute significantly, and will need to be investigated.  

\citet{kaviraj10} also show that the majority of dET have blue $NUV-r$ rest-frame colours, 
a signature of episodes of recent star formation (RSF). The fraction of the stellar mass formed in the 
RSF is $f_{\star, RSF} \sim 3$\%-20\% (see also \citealt{scarlata07ee,kaviraj08}), while the derived metallicity 
makes unlikely gas-rich mergers as the origin of this RSF. We find that $\sim$80\% of the companions of the red 
primaries are blue indicating that there is a gas supply to the RSF, while the stellar mass is dominated by 
the red, old component of both galaxies. Using the recipe provided by \citet{stewart09} to determine 
$M_{\rm gas}/M_{\star}$, where $M_{\rm gas}$ is the mass of gas in the galaxy, we explore the mass and $\mu$ range of our red pairs, and estimate that the gaseous mass is
typically $\lesssim25$\% of the total stellar mass in our red pairs. 
Simulations suggest that $\sim50-75$\% 
of the gas in mergers can be consumed to form new stars \citep{cox04,cox06gas}. This leads to a 
$f_{\star, RSF} \lesssim 20$\%, in agreement with the observed mass formed in the RSF episodes. This result 
is supporting mergers as the main cause of RSF in early-type galaxies since $z \sim 1$ \citep[see also][]{onti11}.

\citet{bundy10} find that the red sequence is populated not only by E/S0 galaxies, but also by passive, 
early-type (i.e., bulge dominated) spirals. While $80$\% of the mergers experienced by a red galaxy are 
with a blue SF companion, the low gaseous mass involved in these mergers ($\lesssim25$\%) prevent the 
regrowth of a spiral disc \citep{hopkins09disk}. Hence, our observed merger rate could be enough to transform 
the red, early-type spirals into E/S0 galaxies. A more detailed study of the merger fraction of red galaxies as 
a function of their morphology is needed to understand the transition between red spirals and E/S0 galaxies.

Summarizing, our measured merger rates of bright red galaxies are in agreement with the mass and size 
evolution of massive red galaxies since $z = 1$, and with the frequency of distorted early-type galaxies at $z\sim 0.6$. Minor mergers have a significant impact in the evolution of these massive red galaxies, accounting of $\sim$20\% of the observed evolution.

\subsection{The role of minor mergers in the evolution of blue galaxies}\label{bluemerger}
Observations and $N$-body simulations suggest that major mergers between gas-rich late-type galaxies are an 
efficient way to obtain quiescent, early-type galaxies 
\citep{naab06ss,rothberg06a,rothberg06b,rothberg10,hopkins08ss,hopkins09disk}. 
Recent studies find that gas-rich major mergers can only account for 20\%-30\% of the number density 
evolution in the red sequence of intermediate-mass 
($M_{\star} \gtrsim 10^{10}\ M_{\odot}$) galaxies since $z = 1$ \citep{bundy09,wild09,deravel09,clsj10megoods,clsj10pargoods}, 
while major mergers are enough to explain the number evolution of massive galaxies in the same redshift range
($M_{\star} \gtrsim 10^{11}\ M_{\odot}$, \citealt{eliche10I,eliche10II,robaina10,oesch10}).

Hence, we need other mechanisms than major mergers to transform intermediate-mass blue cloud galaxies 
into red sequence ones. One possible mechanism is minor merging. The $N$--body simulations find that 
minor mergers increase the S\'ersic index of galaxies \citep{eliche06} and that several minor mergers 
have the same effect as a major one: only the final mass accreted is important (i.e., ten 1/10 mergers 
are equivalent to one equal-mass merger, \citealt{bournaud07}). We find that the minor-to-major merger ratio 
of blue galaxies increases between $z = 0.8$ and $z = 0.5$ from $\sim0.5$ to $\sim2$, indicating that minor mergers may  
play an important role in the growth of the red sequence since $z \sim 0.5$. However, we find that the mass accreted by minor 
mergers is $\sim15$\% of the mass accreted by major mergers at $z = 0.8$, and $\sim 0.6$ at $z = 0.5$.  
Even in the lower redshift range, where minor mergers are twice more common than major ones in blue galaxies, the latter are more 
efficient in transforming gas-rich galaxies into E/S0. In addition, the observed $R_{\rm mm}^{\rm blue}$ implies that, 
in the range $[0.2,0.95)$, a gas-rich galaxy have only undergone $N_{\rm mm}^{\rm blue}\sim 0.15$ minor mergers, 
making it unlikely that a gas-rich galaxy suffers more than one minor merger since $z \sim 1$. 
In summary, our observations indicate that minor mergers 
affect less the structure of gas-rich galaxies than major mergers in the redshift range under study, and they can lead into 
early spirals instead of into E/S0.

It is also expected that secular processes can transform late spirals into early ones. Bars and disk instabilities 
support the growth of the central part of the galaxies, called pseudo-bulges \citep{kormendy04,fisher09}. The similar disc and nuclear colours of spirals up to $z \sim 0.8$ \citep{palmero08} also points towards a coordinated growth 
of the bulge and the disc, while \citet{masters10}, and \citet{sheth08} and \citet{cameron10} find that early-type spirals have higher bar fractions
than late-type ones in Galaxy Zoo ($z \sim 0.04$), and COSMOS\footnote{Cosmological Evolution Survey, \citealt{scoville07} (http://cosmos.astro.caltech.edu/index.html).} ($0.2 < z < 0.85$), respectively. The comparison of the observational \citep[this paper,][]{clsj10pargoods} and theoretical \citep{oesch10} major + minor 
merger rate against the number density growth of intermediate-mass, early-type galaxies also suggests that secular 
processes are needed. 

If these early, bulge-dominated systems, whatever their origin, have their star formation shut down by some processes 
unrelated to mergers, as gas exhaustion \citep{zheng07,bauer10} or some form of quenching (e.g., morphological quenching, \citealt{martig09}; or environment quenching, \citealt{peng10}), 
they then become passive early-type disc on the red sequence, as those found by \citet{bundy10}.

It is also worth noting that because the merger fraction increases when $\mu$ decreases, it is possible that galaxies smaller / fainter than
studied in this paper may play a significant role. However, we find that the increase in the merger fraction cannot compensate 
for the decrease in the mass of the companion and the increase in the typical merger timescale, so it is not expected that 
mergers with $\mu < 1/10$ have been important in the evolution of intermediate-mass gas-rich galaxies. Cosmological models also suggest that merger events lower than $\mu < 1/10$ have little impact (less than $10$\%) in the mass assembly of spheroids 
\citep{hopkins10fusbul}.

\section{Summary and conclusions}\label{conclusion}
We have estimated, for the first time in the literature, the minor merger fraction and rate of $L_{B} \gtrsim L_{B}^{*}$ galaxies from kinematically confirmed close pairs, reaching the minor companion regime, $1/10 \leq \mu < 1/4$ ($\Delta M_{B} = 1.5 - 2.5$) thanks to the deep spectroscopy in VVDS-Deep ($I_{\rm AB} \leq 24$), and robust statistics in a wide 0.5 deg$^{2}$ area.

We find that minor mergers for bright galaxies show little evolution with redshift as a power-law $(1+z)^m$ with index $m=-0.4\pm0.7$ for the merger fraction and $m=-0.5\pm0.7$ for the merger rate, while the major merger fraction ($m = 1.3\pm0.5$) and rate ($m = 1.3\pm0.6$) for the same galaxies increases. The dependence of the merger fraction on $\mu$ is well described  by a power-law function, $f_{m}\,(\geq\,\mu) \propto \mu^{s}$. The value of $s$ for the complete magnitude-limited sample, $M_{B}^{\rm e} \leq -20$, evolves from $s = -0.60\pm0.08$ at $z = 0.8$ to $s = -1.02\pm0.13$ at $z = 0.5$. 
When we split our bright galaxies in red and blue following the rest-colour bimodality, we find that in the redshift range explored i) $f_{\rm m}$ is higher for red galaxies at every $\mu$, ii) $f_{\rm m}^{\rm red}$ does not evolve with $z$, with $s = -0.79\pm0.12$ at $0.2 < z < 0.95$, and iii) $f_{\rm m}^{\rm blue}$ evolves dramatically: the major merger fraction of blue galaxies decreases by a factor of three with cosmic time, while the minor merger fraction of blue galaxies is roughly constant.

Our results show that normal $L_B \gtrsim L_{B}^{*}$ galaxies have undergone 0.4 minor and 0.4 major mergers since $z \sim 1$, which implies a total mass growth from major and minor mergers with $\mu \geq 1/10$ by about 25\%. The relative contribution of the mass growth by merging is $\sim 25$\% due to minor mergers with $1/10 \leq \mu < 1/4$ and $\sim$75\% due to major mergers with $ \mu \geq 1/4$. The relative effect of merging is more important for red than for blue galaxies, with red galaxies subject to 0.5 minor and 0.7 major mergers since $z\sim1$. This leads to a mass growth of $\sim40$\% and a size increase by a factor of 2 of red galaxies, in agreement with the evolution of massive galaxies as reported by previous works \cite[e.g.,][]{vanderwel08,vandokkum10}. This supports that mergers are an important contributor to the evolution of the most massive red galaxies since $z \sim 1$. For blue galaxies, our results imply that minor mergers likely lead to early-type spirals rather than elliptical galaxies. 

Our analysis therefore shows that minor merging is a significant but not dominant mechanism contributing to the mass growth of galaxies in the last $\sim 8$ Gyr. Merging alone is not sufficient to explain the observed mass growth of galaxies, and other processes must therefore be operating. The contribution from minor merging of low mass companions with $\mu < 1/10$ is yet to be estimated, but we expect that this contribution would have only limited effects. 

To expand on our observational results, the study of the minor merger fraction in other fields will be needed to minimize cosmic variance effect, on larger samples to better constrain the evolution of $f_{\rm mm}$ with redshift. In addition, the study of the  dependence of minor mergers on properties like mass, morphology or environment will provide other important clues about the role of mergers in the evolution of galaxies since $z \sim 1$. It is also worth noting that direct measurements of the minor merger fraction have yet to be secured at low redshift, while these will be needed to better constrain the minor merger fraction evolution with $z$.

\begin{acknowledgements}
We dedicate this paper to the memory of our six IAC colleagues and friends who
met with a fatal accident in Piedra de los Cochinos, Tenerife, in February 2007,
with a special thanks to Maurizio Panniello, whose teachings of \texttt{python}
were so important for this paper. 

We thank the anonymous referee for his/her comments and suggestions, that improved
the quality of the paper.

C. L. S. acknowledge the funding support of ANR-07-BLAN-0228 and the help of 
A. Ealet in the statistical analysis. 

A. P. has beed supported by the research grant of the Polish Ministry of
Science Nr N N203 51 29 38 and the European Associated Laboratory
"Astrophysics Poland-France".

This work uses the Millennium Galaxy Catalogue, 
which consists of imaging data from the
Isaac Newton Telescope and spectroscopic data from the Anglo
Australian Telescope, the ANU 2.3m, the ESO New Technology Telescope,
the Telescopio Nazionale Galileo, and the Gemini North Telescope. This
survey was supported through grants from the Particle Physics and
Astronomy Research Council (UK) and the Australian Research Council
(AUS). The data and data products of this survey are publicly available from
http://www.eso.org/~jliske/mgc/ or on request from J. Liske or
S.P. Driver.
\end{acknowledgements}

\bibliography{biblio}

\begin{thebibliography}{132}
\expandafter\ifx\csname natexlab\endcsname\relax\def\natexlab#1{#1}\fi

\bibitem[{{Abbas} {et~al.}(2010){Abbas}, {de La Torre}, {Le F{\`e}vre},
  {Guzzo}, {Marinoni}, {Meneux}, {Pollo}, {Zamorani}, {Bottini}, {Garilli}, {Le
  Brun}, {Maccagni}, {Scaramella}, {Scodeggio}, {Tresse}, {Vettolani},
  {Zanichelli}, {Adami}, {Arnouts}, {Bardelli}, {Bolzonella}, {Cappi},
  {Charlot}, {Ciliegi}, {Contini}, {Foucaud}, {Franzetti}, {Gavignaud},
  {Ilbert}, {Iovino}, {Lamareille}, {McCracken}, {Marano}, {Mazure}, {Merighi},
  {Paltani}, {Pell{\`o}}, {Pozzetti}, {Radovich}, {Vergani}, {Zucca}, {Bondi},
  {Bongiorno}, {Brinchmann}, {Cucciati}, {de Ravel}, {Gregorini},
  {Perez-Montero}, {Mellier}, \& {Merluzzi}}]{abbas10}
{Abbas}, U., {de La Torre}, S., {Le F{\`e}vre}, O., {et~al.} 2010, \mnras, 406,
  1306

\bibitem[{{Adelman-McCarthy} {et~al.}(2006){Adelman-McCarthy}, {Ag{\"u}eros},
  {Allam}, {Anderson}, {Anderson}, {Annis}, {Bahcall}, {Baldry}, {Barentine},
  {Berlind}, {Bernardi}, {Blanton}, {Boroski}, {Brewington}, {Brinchmann},
  {Brinkmann}, {Brunner}, {Budav{\'a}ri}, {Carey}, {Carr}, {Castander},
  {Connolly}, {Csabai}, {Czarapata}, {Dalcanton}, {Doi}, {Dong}, {Eisenstein},
  {Evans}, {Fan}, {Finkbeiner}, {Friedman}, {Frieman}, {Fukugita}, {Gillespie},
  {Glazebrook}, {Gray}, {Grebel}, {Gunn}, {Gurbani}, {de Haas}, {Hall},
  {Harris}, {Harvanek}, {Hawley}, {Hayes}, {Hendry}, {Hennessy}, {Hindsley},
  {Hirata}, {Hogan}, {Hogg}, {Holmgren}, {Holtzman}, {Ichikawa}, {Ivezi{\'c}},
  {Jester}, {Johnston}, {Jorgensen}, {Juri{\'c}}, {Kent}, {Kleinman}, {Knapp},
  {Kniazev}, {Kron}, {Krzesinski}, {Kuropatkin}, {Lamb}, {Lampeitl}, {Lee},
  {Leger}, {Lin}, {Long}, {Loveday}, {Lupton}, {Margon},
  {Mart{\'{\i}}nez-Delgado}, {Mandelbaum}, {Matsubara}, {McGehee}, {McKay},
  {Meiksin}, {Munn}, {Nakajima}, {Nash}, {Neilsen}, {Newberg}, {Newman},
  {Nichol}, {Nicinski}, {Nieto-Santisteban}, {Nitta}, {O'Mullane}, {Okamura},
  {Owen}, {Padmanabhan}, {Pauls}, {Peoples}, {Pier}, {Pope}, {Pourbaix},
  {Quinn}, {Richards}, {Richmond}, {Rockosi}, {Schlegel}, {Schneider},
  {Schroeder}, {Scranton}, {Seljak}, {Sheldon}, {Shimasaku}, {Smith}, {Smol{\v
  c}i{\'c}}, {Snedden}, {Stoughton}, {Strauss}, {SubbaRao}, {Szalay},
  {Szapudi}, {Szkody}, {Tegmark}, {Thakar}, {Tucker}, {Uomoto}, {Vanden Berk},
  {Vandenberg}, {Vogeley}, {Voges}, {Vogt}, {Walkowicz}, {Weinberg}, {West},
  {White}, {Xu}, {Yanny}, {Yocum}, {York}, {Zehavi}, {Zibetti}, \&
  {Zucker}}]{adelman06}
{Adelman-McCarthy}, J.~K., {Ag{\"u}eros}, M.~A., {Allam}, S.~S., {et~al.} 2006,
  \apjs, 162, 38

\bibitem[{{Aitken}(1935)}]{aitken34}
{Aitken}, A.~C. 1935, Proc. R. Soc. Edinb, 55, 42

\bibitem[{{Arnouts} {et~al.}(2007){Arnouts}, {Walcher}, {Le F{\`e}vre},
  {Zamorani}, {Ilbert}, {Le Brun}, {Pozzetti}, {Bardelli}, {Tresse}, {Zucca},
  {Charlot}, {Lamareille}, {McCracken}, {Bolzonella}, {Iovino}, {Lonsdale},
  {Polletta}, {Surace}, {Bottini}, {Garilli}, {Maccagni}, {Picat},
  {Scaramella}, {Scodeggio}, {Vettolani}, {Zanichelli}, {Adami}, {Cappi},
  {Ciliegi}, {Contini}, {de La Torre}, {Foucaud}, {Franzetti}, {Gavignaud},
  {Guzzo}, {Marano}, {Marinoni}, {Mazure}, {Meneux}, {Merighi}, {Paltani},
  {Pell{\`o}}, {Pollo}, {Radovich}, {Temporin}, \& {Vergani}}]{arnouts07}
{Arnouts}, S., {Walcher}, C.~J., {Le F{\`e}vre}, O., {et~al.} 2007, \aap, 476,
  137

\bibitem[{{Bauermeister} {et~al.}(2010){Bauermeister}, {Blitz}, \&
  {Ma}}]{bauer10}
{Bauermeister}, A., {Blitz}, L., \& {Ma}, C. 2010, \apj, 717, 323

\bibitem[{{Bell} {et~al.}(2006){Bell}, {Phleps}, {Somerville}, {Wolf}, {Borch},
  \& {Meisenheimer}}]{bell06}
{Bell}, E.~F., {Phleps}, S., {Somerville}, R.~S., {et~al.} 2006, \apj, 652, 270

\bibitem[{{Bezanson} {et~al.}(2009){Bezanson}, {van Dokkum}, {Tal},
  {Marchesini}, {Kriek}, {Franx}, \& {Coppi}}]{bezanson09}
{Bezanson}, R., {van Dokkum}, P.~G., {Tal}, T., {et~al.} 2009, \apj, 697, 1290

\bibitem[{{Bournaud} {et~al.}(2007){Bournaud}, {Jog}, \& {Combes}}]{bournaud07}
{Bournaud}, F., {Jog}, C.~J., \& {Combes}, F. 2007, \aap, 476, 1179

\bibitem[{{Brammer} {et~al.}(2009){Brammer}, {Whitaker}, {van Dokkum},
  {Marchesini}, {Labb{\'e}}, {Franx}, {Kriek}, {Quadri}, {Illingworth}, {Lee},
  {Muzzin}, \& {Rudnick}}]{brammer09}
{Brammer}, G.~B., {Whitaker}, K.~E., {van Dokkum}, P.~G., {et~al.} 2009, \apjl,
  706, L173

\bibitem[{{Bridge} {et~al.}(2010){Bridge}, {Carlberg}, \&
  {Sullivan}}]{bridge10}
{Bridge}, C.~R., {Carlberg}, R.~G., \& {Sullivan}, M. 2010, \apj, 709, 1067

\bibitem[{{Brown} {et~al.}(2007){Brown}, {Dey}, {Jannuzi}, {Brand}, {Benson},
  {Brodwin}, {Croton}, \& {Eisenhardt}}]{brown07}
{Brown}, M.~J.~I., {Dey}, A., {Jannuzi}, B.~T., {et~al.} 2007, \apj, 654, 858

\bibitem[{{Brown} {et~al.}(2008){Brown}, {Zheng}, {White}, {Dey}, {Jannuzi},
  {Benson}, {Brand}, {Brodwin}, \& {Croton}}]{brown08}
{Brown}, M.~J.~I., {Zheng}, Z., {White}, M., {et~al.} 2008, \apj, 682, 937

\bibitem[{{Buitrago} {et~al.}(2008){Buitrago}, {Trujillo}, {Conselice},
  {Bouwens}, {Dickinson}, \& {Yan}}]{buitrago08}
{Buitrago}, F., {Trujillo}, I., {Conselice}, C.~J., {et~al.} 2008, \apjl, 687,
  L61

\bibitem[{{Bundy} {et~al.}(2005){Bundy}, {Ellis}, \& {Conselice}}]{bundy05}
{Bundy}, K., {Ellis}, R.~S., \& {Conselice}, C.~J. 2005, \apj, 625, 621

\bibitem[{{Bundy} {et~al.}(2009){Bundy}, {Fukugita}, {Ellis}, {Targett},
  {Belli}, \& {Kodama}}]{bundy09}
{Bundy}, K., {Fukugita}, M., {Ellis}, R.~S., {et~al.} 2009, \apj, 697, 1369

\bibitem[{{Bundy} {et~al.}(2010){Bundy}, {Scarlata}, {Carollo}, {Ellis},
  {Drory}, {Hopkins}, {Salvato}, {Leauthaud}, {Koekemoer}, {Murray}, {Ilbert},
  {Oesch}, {Ma}, {Capak}, {Pozzetti}, \& {Scoville}}]{bundy10}
{Bundy}, K., {Scarlata}, C., {Carollo}, C.~M., {et~al.} 2010, \apj, 719, 1969

\bibitem[{{Cameron}(2010)}]{cameron10sig}
{Cameron}, E. 2010, PASA, submitted [ArXiv:1012.0566]

\bibitem[{{Cameron} {et~al.}(2010){Cameron}, {Carollo}, {Oesch}, {Aller},
  {Bschorr}, {Cerulo}, {Aussel}, {Capak}, {Le Floc'h}, {Ilbert}, {Kneib},
  {Koekemoer}, {Leauthaud}, {Lilly}, {Massey}, {McCracken}, {Rhodes},
  {Salvato}, {Sanders}, {Scoville}, {Sheth}, {Taniguchi}, \&
  {Thompson}}]{cameron10}
{Cameron}, E., {Carollo}, C.~M., {Oesch}, P., {et~al.} 2010, \mnras, 409, 346

\bibitem[{{Conselice}(2006)}]{conselice06ff}
{Conselice}, C.~J. 2006, \apj, 638, 686

\bibitem[{{Conselice} {et~al.}(2003){Conselice}, {Bershady}, {Dickinson}, \&
  {Papovich}}]{conselice03ff}
{Conselice}, C.~J., {Bershady}, M.~A., {Dickinson}, M., \& {Papovich}, C. 2003,
  \aj, 126, 1183

\bibitem[{{Conselice} {et~al.}(2007){Conselice}, {Bundy}, {Trujillo}, {Coil},
  {Eisenhardt}, {Ellis}, {Georgakakis}, {Huang}, {Lotz}, {Nandra}, {Newman},
  {Papovich}, {Weiner}, \& {Willmer}}]{conselice07}
{Conselice}, C.~J., {Bundy}, K., {Trujillo}, I., {et~al.} 2007, \mnras, 381,
  962

\bibitem[{{Conselice} {et~al.}(2009){Conselice}, {Yang}, \&
  {Bluck}}]{conselice09cos}
{Conselice}, C.~J., {Yang}, C., \& {Bluck}, A.~F.~L. 2009, \mnras, 361

\bibitem[{{Cool} {et~al.}(2008){Cool}, {Eisenstein}, {Fan}, {Fukugita},
  {Jiang}, {Maraston}, {Meiksin}, {Schneider}, \& {Wake}}]{cool08}
{Cool}, R.~J., {Eisenstein}, D.~J., {Fan}, X., {et~al.} 2008, \apj, 682, 919

\bibitem[{{Cortese} {et~al.}(2008){Cortese}, {Boselli}, {Franzetti}, {Decarli},
  {Gavazzi}, {Boissier}, \& {Buat}}]{cortese08}
{Cortese}, L., {Boselli}, A., {Franzetti}, P., {et~al.} 2008, \mnras, 386, 1157

\bibitem[{{Cox} {et~al.}(2006){Cox}, {Jonsson}, {Primack}, \&
  {Somerville}}]{cox06gas}
{Cox}, T.~J., {Jonsson}, P., {Primack}, J.~R., \& {Somerville}, R.~S. 2006,
  \mnras, 373, 1013

\bibitem[{{Cox} {et~al.}(2004){Cox}, {Primack}, {Jonsson}, \&
  {Somerville}}]{cox04}
{Cox}, T.~J., {Primack}, J., {Jonsson}, P., \& {Somerville}, R.~S. 2004, \apjl,
  607, L87

\bibitem[{{Cucciati} {et~al.}(2010){Cucciati}, {Marinoni}, {Iovino},
  {Bardelli}, {Adami}, {Mazure}, {Scodeggio}, {Maccagni}, {Temporin}, {Zucca},
  {De Lucia}, {Blaizot}, {Garilli}, {Meneux}, {Zamorani}, {Le F{\`e}vre},
  {Cappi}, {Guzzo}, {Bottini}, {Le Brun}, {Tresse}, {Vettolani}, {Zanichelli},
  {Arnouts}, {Bolzonella}, {Charlot}, {Ciliegi}, {Contini}, {Foucaud},
  {Franzetti}, {Gavignaud}, {Ilbert}, {Lamareille}, {McCracken}, {Marano},
  {Merighi}, {Paltani}, {Pell{\`o}}, {Pollo}, {Pozzetti}, {Vergani}, \&
  {P{\'e}rez-Montero}}]{cucciati10}
{Cucciati}, O., {Marinoni}, C., {Iovino}, A., {et~al.} 2010, \aap, 520, A42+

\bibitem[{{Daddi} {et~al.}(2005){Daddi}, {Renzini}, {Pirzkal}, {Cimatti},
  {Malhotra}, {Stiavelli}, {Xu}, {Pasquali}, {Rhoads}, {Brusa}, {di Serego
  Alighieri}, {Ferguson}, {Koekemoer}, {Moustakas}, {Panagia}, \&
  {Windhorst}}]{daddi05}
{Daddi}, E., {Renzini}, A., {Pirzkal}, N., {et~al.} 2005, \apj, 626, 680

\bibitem[{{Darg} {et~al.}(2010){Darg}, {Kaviraj}, {Lintott}, {Schawinski},
  {Sarzi}, {Bamford}, {Silk}, {Proctor}, {Andreescu}, {Murray}, {Nichol},
  {Raddick}, {Slosar}, {Szalay}, {Thomas}, \& {Vandenberg}}]{darg10i}
{Darg}, D.~W., {Kaviraj}, S., {Lintott}, C.~J., {et~al.} 2010, \mnras, 401,
  1043

\bibitem[{{De Propris} {et~al.}(2007){De Propris}, {Conselice}, {Liske},
  {Driver}, {Patton}, {Graham}, \& {Allen}}]{depropris07}
{De Propris}, R., {Conselice}, C.~J., {Liske}, J., {et~al.} 2007, \apj, 666,
  212

\bibitem[{{De Propris} {et~al.}(2005){De Propris}, {Liske}, {Driver}, {Allen},
  \& {Cross}}]{depropris05}
{De Propris}, R., {Liske}, J., {Driver}, S.~P., {Allen}, P.~D., \& {Cross},
  N.~J.~G. 2005, \aj, 130, 1516

\bibitem[{{de Ravel} {et~al.}(2009){de Ravel}, {Le F{\`e}vre}, {Tresse},
  {Bottini}, {Garilli}, {Le Brun}, {Maccagni}, {Scaramella}, {Scodeggio},
  {Vettolani}, {Zanichelli}, {Adami}, {Arnouts}, {Bardelli}, {Bolzonella},
  {Cappi}, {Charlot}, {Ciliegi}, {Contini}, {Foucaud}, {Franzetti},
  {Gavignaud}, {Guzzo}, {Ilbert}, {Iovino}, {Lamareille}, {McCracken},
  {Marano}, {Marinoni}, {Mazure}, {Meneux}, {Merighi}, {Paltani}, {Pell{\`o}},
  {Pollo}, {Pozzetti}, {Radovich}, {Vergani}, {Zamorani}, {Zucca}, {Bondi},
  {Bongiorno}, {Brinchmann}, {Cucciati}, {de La Torre}, {Gregorini}, {Memeo},
  {Perez-Montero}, {Mellier}, {Merluzzi}, \& {Temporin}}]{deravel09}
{de Ravel}, L., {Le F{\`e}vre}, O., {Tresse}, L., {et~al.} 2009, \aap, 498, 379

\bibitem[{{Dom{\'{\i}}nguez-Palmero} \& {Balcells}(2008)}]{palmero08}
{Dom{\'{\i}}nguez-Palmero}, L. \& {Balcells}, M. 2008, \aap, 489, 1003

\bibitem[{{Driver} {et~al.}(2005){Driver}, {Liske}, {Cross}, {De Propris}, \&
  {Allen}}]{driver05}
{Driver}, S.~P., {Liske}, J., {Cross}, N.~J.~G., {De Propris}, R., \& {Allen},
  P.~D. 2005, \mnras, 360, 81

\bibitem[{{Drory} \& {Alvarez}(2008)}]{drory08}
{Drory}, N. \& {Alvarez}, M. 2008, \apj, 680, 41

\bibitem[{{Drory} {et~al.}(2005){Drory}, {Salvato}, {Gabasch}, {Bender},
  {Hopp}, {Feulner}, \& {Pannella}}]{drory05}
{Drory}, N., {Salvato}, M., {Gabasch}, A., {et~al.} 2005, \apjl, 619, L131

\bibitem[{{Efron}(1982)}]{efron82}
{Efron}, B. 1982

\bibitem[{{Eliche-Moral} {et~al.}(2006){Eliche-Moral}, {Balcells}, {Aguerri},
  \& {Gonz{\'a}lez-Garc{\'{\i}}a}}]{eliche06}
{Eliche-Moral}, M.~C., {Balcells}, M., {Aguerri}, J.~A.~L., \&
  {Gonz{\'a}lez-Garc{\'{\i}}a}, A.~C. 2006, \aap, 457, 91

\bibitem[{{Eliche-Moral} {et~al.}(2010{\natexlab{a}}){Eliche-Moral}, {Prieto},
  {Gallego}, {Barro}, {Zamorano}, {Lopez-Sanjuan}, {Balcells}, {Guzman}, \&
  {Munoz-Mateos}}]{eliche10I}
{Eliche-Moral}, M.~C., {Prieto}, M., {Gallego}, J., {et~al.}
  2010{\natexlab{a}}, \aap, 519, A55

\bibitem[{{Eliche-Moral} {et~al.}(2010{\natexlab{b}}){Eliche-Moral}, {Prieto},
  {Gallego}, \& {Zamorano}}]{eliche10II}
{Eliche-Moral}, M.~C., {Prieto}, M., {Gallego}, J., \& {Zamorano}, J.
  2010{\natexlab{b}}, \apj, submitted [ArXiv: 1003.0686]

\bibitem[{{Ellison} {et~al.}(2008){Ellison}, {Patton}, {Simard}, \&
  {McConnachie}}]{ellison08}
{Ellison}, S.~L., {Patton}, D.~R., {Simard}, L., \& {McConnachie}, A.~W. 2008,
  \aj, 135, 1877

\bibitem[{{Faber} {et~al.}(2007){Faber}, {Willmer}, {Wolf}, {Koo}, {Weiner},
  {Newman}, {Im}, {Coil}, {Conroy}, {Cooper}, {Davis}, {Finkbeiner}, {Gerke},
  {Gebhardt}, {Groth}, {Guhathakurta}, {Harker}, {Kaiser}, {Kassin},
  {Kleinheinrich}, {Konidaris}, {Kron}, {Lin}, {Luppino}, {Madgwick},
  {Meisenheimer}, {Noeske}, {Phillips}, {Sarajedini}, {Schiavon}, {Simard},
  {Szalay}, {Vogt}, \& {Yan}}]{faber07}
{Faber}, S.~M., {Willmer}, C.~N.~A., {Wolf}, C., {et~al.} 2007, \apj, 665, 265

\bibitem[{{Fern{\'a}ndez-Ontiveros} {et~al.}(2011){Fern{\'a}ndez-Ontiveros},
  {L{\'o}pez-Sanjuan}, {Montes}, {Prieto}, \& {Acosta-Pulido}}]{onti11}
{Fern{\'a}ndez-Ontiveros}, J.~A., {L{\'o}pez-Sanjuan}, C., {Montes}, M.,
  {Prieto}, M.~A., \& {Acosta-Pulido}, J.~A. 2011, \mnras, 411, L21

\bibitem[{{Fisher} {et~al.}(2009){Fisher}, {Drory}, \& {Fabricius}}]{fisher09}
{Fisher}, D.~B., {Drory}, N., \& {Fabricius}, M.~H. 2009, \apj, 697, 630

\bibitem[{{Fontana} {et~al.}(2009){Fontana}, {Santini}, {Grazian},
  {Pentericci}, {Fiore}, {Castellano}, {Giallongo}, {Menci}, {Salimbeni},
  {Cristiani}, {Nonino}, \& {Vanzella}}]{fontana09}
{Fontana}, A., {Santini}, P., {Grazian}, A., {et~al.} 2009, \aap, 501, 15

\bibitem[{{Franzetti} {et~al.}(2007){Franzetti}, {Scodeggio}, {Garilli},
  {Vergani}, {Maccagni}, {Guzzo}, {Tresse}, {Ilbert}, {Lamareille}, {Contini},
  {Le F{\`e}vre}, {Zamorani}, {Brinchmann}, {Charlot}, {Bottini}, {Le Brun},
  {Picat}, {Scaramella}, {Vettolani}, {Zanichelli}, {Adami}, {Arnouts},
  {Bardelli}, {Bolzonella}, {Cappi}, {Ciliegi}, {Foucaud}, {Gavignaud},
  {Iovino}, {McCracken}, {Marano}, {Marinoni}, {Mazure}, {Meneux}, {Merighi},
  {Paltani}, {Pell{\`o}}, {Pollo}, {Pozzetti}, {Radovich}, {Zucca}, {Cucciati},
  \& {Walcher}}]{franzetti07}
{Franzetti}, P., {Scodeggio}, M., {Garilli}, B., {et~al.} 2007, \aap, 465, 711

\bibitem[{{Genel} {et~al.}(2010){Genel}, {Bouch{\'e}}, {Naab}, {Sternberg}, \&
  {Genzel}}]{genel10}
{Genel}, S., {Bouch{\'e}}, N., {Naab}, T., {Sternberg}, A., \& {Genzel}, R.
  2010, \apj, 719, 229

\bibitem[{{Giallongo} {et~al.}(2005){Giallongo}, {Salimbeni}, {Menci},
  {Zamorani}, {Fontana}, {Dickinson}, {Cristiani}, \& {Pozzetti}}]{giallongo05}
{Giallongo}, E., {Salimbeni}, S., {Menci}, N., {et~al.} 2005, \apj, 622, 116

\bibitem[{{Gonz{\'a}lez-Garc{\'{\i}}a}
  {et~al.}(2009){Gonz{\'a}lez-Garc{\'{\i}}a}, {O{\~n}orbe},
  {Dom{\'{\i}}nguez-Tenreiro}, \& {G{\'o}mez-Flechoso}}]{gongar09}
{Gonz{\'a}lez-Garc{\'{\i}}a}, A.~C., {O{\~n}orbe}, J.,
  {Dom{\'{\i}}nguez-Tenreiro}, R., \& {G{\'o}mez-Flechoso}, M.~{\'A}. 2009,
  \aap, 497, 35

\bibitem[{{Hopkins} {et~al.}(2010{\natexlab{a}}){Hopkins}, {Bundy}, {Croton},
  {Hernquist}, {Keres}, {Khochfar}, {Stewart}, {Wetzel}, \&
  {Younger}}]{hopkins10fusbul}
{Hopkins}, P.~F., {Bundy}, K., {Croton}, D., {et~al.} 2010{\natexlab{a}}, \apj,
  715, 202

\bibitem[{{Hopkins} {et~al.}(2010{\natexlab{b}}){Hopkins}, {Bundy},
  {Hernquist}, {Wuyts}, \& {Cox}}]{hopkins10size}
{Hopkins}, P.~F., {Bundy}, K., {Hernquist}, L., {Wuyts}, S., \& {Cox}, T.~J.
  2010{\natexlab{b}}, \mnras, 401, 1099

\bibitem[{{Hopkins} {et~al.}(2009{\natexlab{a}}){Hopkins}, {Bundy}, {Murray},
  {Quataert}, {Lauer}, \& {Ma}}]{hopkins09core}
{Hopkins}, P.~F., {Bundy}, K., {Murray}, N., {et~al.} 2009{\natexlab{a}},
  \mnras, 398, 898

\bibitem[{{Hopkins} {et~al.}(2009{\natexlab{b}}){Hopkins}, {Cox}, {Younger}, \&
  {Hernquist}}]{hopkins09disk}
{Hopkins}, P.~F., {Cox}, T.~J., {Younger}, J.~D., \& {Hernquist}, L.
  2009{\natexlab{b}}, \apj, 691, 1168

\bibitem[{{Hopkins} {et~al.}(2008){Hopkins}, {Hernquist}, {Cox}, {Dutta}, \&
  {Rothberg}}]{hopkins08ss}
{Hopkins}, P.~F., {Hernquist}, L., {Cox}, T.~J., {Dutta}, S.~N., \& {Rothberg},
  B. 2008, \apj, 679, 156

\bibitem[{{Hopkins} {et~al.}(2009{\natexlab{c}}){Hopkins}, {Somerville}, {Cox},
  {Hernquist}, {Jogee}, {Kere{\v s}}, {Ma}, {Robertson}, \&
  {Stewart}}]{hopkins09bulges}
{Hopkins}, P.~F., {Somerville}, R.~S., {Cox}, T.~J., {et~al.}
  2009{\natexlab{c}}, \mnras, 397, 802

\bibitem[{{Ilbert} {et~al.}(2006){Ilbert}, {Arnouts}, {McCracken},
  {Bolzonella}, {Bertin}, {Le F{\`e}vre}, {Mellier}, {Zamorani}, {Pell{\`o}},
  {Iovino}, {Tresse}, {Le Brun}, {Bottini}, {Garilli}, {Maccagni}, {Picat},
  {Scaramella}, {Scodeggio}, {Vettolani}, {Zanichelli}, {Adami}, {Bardelli},
  {Cappi}, {Charlot}, {Ciliegi}, {Contini}, {Cucciati}, {Foucaud}, {Franzetti},
  {Gavignaud}, {Guzzo}, {Marano}, {Marinoni}, {Mazure}, {Meneux}, {Merighi},
  {Paltani}, {Pollo}, {Pozzetti}, {Radovich}, {Zucca}, {Bondi}, {Bongiorno},
  {Busarello}, {de La Torre}, {Gregorini}, {Lamareille}, {Mathez}, {Merluzzi},
  {Ripepi}, {Rizzo}, \& {Vergani}}]{ilbert06phot}
{Ilbert}, O., {Arnouts}, S., {McCracken}, H.~J., {et~al.} 2006, \aap, 457, 841

\bibitem[{{Ilbert} {et~al.}(2010){Ilbert}, {Salvato}, {Le Floc'h}, {Aussel},
  {Capak}, {McCracken}, {Mobasher}, {Kartaltepe}, {Scoville}, {Sanders},
  {Arnouts}, {Bundy}, {Cassata}, {Kneib}, {Koekemoer}, {Le F{\`e}vre}, {Lilly},
  {Surace}, {Taniguchi}, {Tasca}, {Thompson}, {Tresse}, {Zamojski}, {Zamorani},
  \& {Zucca}}]{ilbert10}
{Ilbert}, O., {Salvato}, M., {Le Floc'h}, E., {et~al.} 2010, \apj, 709, 644

\bibitem[{{Ilbert} {et~al.}(2005){Ilbert}, {Tresse}, {Zucca}, {Bardelli},
  {Arnouts}, {Zamorani}, {Pozzetti}, {Bottini}, {Garilli}, {Le Brun}, {Le
  F{\`e}vre}, {Maccagni}, {Picat}, {Scaramella}, {Scodeggio}, {Vettolani},
  {Zanichelli}, {Adami}, {Arnaboldi}, {Bolzonella}, {Cappi}, {Charlot},
  {Contini}, {Foucaud}, {Franzetti}, {Gavignaud}, {Guzzo}, {Iovino},
  {McCracken}, {Marano}, {Marinoni}, {Mathez}, {Mazure}, {Meneux}, {Merighi},
  {Paltani}, {Pello}, {Pollo}, {Radovich}, {Bondi}, {Bongiorno}, {Busarello},
  {Ciliegi}, {Lamareille}, {Mellier}, {Merluzzi}, {Ripepi}, \&
  {Rizzo}}]{ilbert05}
{Ilbert}, O., {Tresse}, L., {Zucca}, E., {et~al.} 2005, \aap, 439, 863

\bibitem[{{Jaech}(1964)}]{jaech64}
{Jaech}, J.~L. 1964, JASA, 59, 863

\bibitem[{{Jiang} {et~al.}(2008){Jiang}, {Jing}, {Faltenbacher}, {Lin}, \&
  {Li}}]{jiang08}
{Jiang}, C.~Y., {Jing}, Y.~P., {Faltenbacher}, A., {Lin}, W.~P., \& {Li}, C.
  2008, \apj

\bibitem[{{Jogee} {et~al.}(2009){Jogee}, {Miller}, {Penner}, {Skelton},
  {Conselice}, {Somerville}, {Bell}, {Zheng}, {Rix}, {Robaina}, {Barazza},
  {Barden}, {Borch}, {Beckwith}, {Caldwell}, {Peng}, {Heymans}, {McIntosh},
  {H{\"a}u{\ss}ler}, {Jahnke}, {Meisenheimer}, {Sanchez}, {Wisotzki}, {Wolf},
  \& {Papovich}}]{jogee09}
{Jogee}, S., {Miller}, S.~H., {Penner}, K., {et~al.} 2009, \apj, 697, 1971

\bibitem[{{Kaviraj} {et~al.}(2008){Kaviraj}, {Khochfar}, {Schawinski}, {Yi},
  {Gawiser}, {Silk}, {Virani}, {Cardamone}, {van Dokkum}, \&
  {Urry}}]{kaviraj08}
{Kaviraj}, S., {Khochfar}, S., {Schawinski}, K., {et~al.} 2008, \mnras, 388, 67

\bibitem[{{Kaviraj} {et~al.}(2009){Kaviraj}, {Peirani}, {Khochfar}, {Silk}, \&
  {Kay}}]{kaviraj09}
{Kaviraj}, S., {Peirani}, S., {Khochfar}, S., {Silk}, J., \& {Kay}, S. 2009,
  \mnras, 394, 1713

\bibitem[{{Kaviraj} {et~al.}(2007){Kaviraj}, {Schawinski}, {Devriendt},
  {Ferreras}, {Khochfar}, {Yoon}, {Yi}, {Deharveng}, {Boselli}, {Barlow},
  {Conrow}, {Forster}, {Friedman}, {Martin}, {Morrissey}, {Neff},
  {Schiminovich}, {Seibert}, {Small}, {Wyder}, {Bianchi}, {Donas}, {Heckman},
  {Lee}, {Madore}, {Milliard}, {Rich}, \& {Szalay}}]{kaviraj07}
{Kaviraj}, S., {Schawinski}, K., {Devriendt}, J.~E.~G., {et~al.} 2007, \apjs,
  173, 619

\bibitem[{{Kaviraj} {et~al.}(2011){Kaviraj}, {Tan}, {Ellis}, \&
  {Silk}}]{kaviraj10}
{Kaviraj}, S., {Tan}, K., {Ellis}, R.~S., \& {Silk}, J. 2011, \mnras, in press
  [ArXiv: 1001.2141]

\bibitem[{{Kitzbichler} \& {White}(2008)}]{kit08}
{Kitzbichler}, M.~G. \& {White}, S.~D.~M. 2008, \mnras, 1300

\bibitem[{{Kormendy} \& {Kennicutt}(2004)}]{kormendy04}
{Kormendy}, J. \& {Kennicutt}, Jr., R.~C. 2004, \araa, 42, 603

\bibitem[{{Le F{\`e}vre} {et~al.}(2000){Le F{\`e}vre}, {Abraham}, {Lilly},
  {Ellis}, {Brinchmann}, {Schade}, {Tresse}, {Colless}, {Crampton},
  {Glazebrook}, {Hammer}, \& {Broadhurst}}]{lefevre00}
{Le F{\`e}vre}, O., {Abraham}, R., {Lilly}, S.~J., {et~al.} 2000, \mnras, 311,
  565

\bibitem[{{Le F{\`e}vre} {et~al.}(2005{\natexlab{a}}){Le F{\`e}vre}, {Guzzo},
  {Meneux}, {Pollo}, {Cappi}, {Colombi}, {Iovino}, {Marinoni}, {McCracken},
  {Scaramella}, {Bottini}, {Garilli}, {Le Brun}, {Maccagni}, {Picat},
  {Scodeggio}, {Tresse}, {Vettolani}, {Zanichelli}, {Adami}, {Arnaboldi},
  {Arnouts}, {Bardelli}, {Blaizot}, {Bolzonella}, {Charlot}, {Ciliegi},
  {Contini}, {Foucaud}, {Franzetti}, {Gavignaud}, {Ilbert}, {Marano}, {Mathez},
  {Mazure}, {Merighi}, {Paltani}, {Pell{\`o}}, {Pozzetti}, {Radovich},
  {Zamorani}, {Zucca}, {Bondi}, {Bongiorno}, {Busarello}, {Lamareille},
  {Mellier}, {Merluzzi}, {Ripepi}, \& {Rizzo}}]{lefevre05cluster}
{Le F{\`e}vre}, O., {Guzzo}, L., {Meneux}, B., {et~al.} 2005{\natexlab{a}},
  \aap, 439, 877

\bibitem[{{Le F{\`e}vre} {et~al.}(2004){Le F{\`e}vre}, {Mellier}, {McCracken},
  {Foucaud}, {Gwyn}, {Radovich}, {Dantel-Fort}, {Bertin}, {Moreau},
  {Cuillandre}, {Pierre}, {Le Brun}, {Mazure}, \& {Tresse}}]{lefevre04img}
{Le F{\`e}vre}, O., {Mellier}, Y., {McCracken}, H.~J., {et~al.} 2004, \aap,
  417, 839

\bibitem[{{Le F{\`e}vre} {et~al.}(2003){Le F{\`e}vre}, {Saisse}, {Mancini},
  {Brau-Nogue}, {Caputi}, {Castinel}, {D'Odorico}, {Garilli}, {Kissler-Patig},
  {Lucuix}, {Mancini}, {Pauget}, {Sciarretta}, {Scodeggio}, {Tresse}, \&
  {Vettolani}}]{lefevre03}
{Le F{\`e}vre}, O., {Saisse}, M., {Mancini}, D., {et~al.} 2003, in Society of
  Photo-Optical Instrumentation Engineers (SPIE) Conference Series, Vol. 4841,
  Society of Photo-Optical Instrumentation Engineers (SPIE) Conference Series,
  ed. {M.~Iye \& A.~F.~M.~Moorwood}, 1670--1681

\bibitem[{{Le F{\`e}vre} {et~al.}(2005{\natexlab{b}}){Le F{\`e}vre},
  {Vettolani}, {Garilli}, {Tresse}, {Bottini}, {Le Brun}, {Maccagni}, {Picat},
  {Scaramella}, {Scodeggio}, {Zanichelli}, {Adami}, {Arnaboldi}, {Arnouts},
  {Bardelli}, {Bolzonella}, {Cappi}, {Charlot}, {Ciliegi}, {Contini},
  {Foucaud}, {Franzetti}, {Gavignaud}, {Guzzo}, {Ilbert}, {Iovino},
  {McCracken}, {Marano}, {Marinoni}, {Mathez}, {Mazure}, {Meneux}, {Merighi},
  {Paltani}, {Pell{\`o}}, {Pollo}, {Pozzetti}, {Radovich}, {Zamorani}, {Zucca},
  {Bondi}, {Bongiorno}, {Busarello}, {Lamareille}, {Mellier}, {Merluzzi},
  {Ripepi}, \& {Rizzo}}]{lefevre05}
{Le F{\`e}vre}, O., {Vettolani}, G., {Garilli}, B., {et~al.}
  2005{\natexlab{b}}, \aap, 439, 845

\bibitem[{{Lin} {et~al.}(2010){Lin}, {Cooper}, {Jian}, {Koo}, {Patton}, {Yan},
  {Willmer}, {Coil}, {Chiueh}, {Croton}, {Gerke}, {Lotz}, {Guhathakurta}, \&
  {Newman}}]{lin10}
{Lin}, L., {Cooper}, M.~C., {Jian}, H., {et~al.} 2010, \apj, 718, 1158

\bibitem[{{Lin} {et~al.}(2004){Lin}, {Koo}, {Willmer}, {Patton}, {Conselice},
  {Yan}, {Coil}, {Cooper}, {Davis}, {Faber}, {Gerke}, {Guhathakurta}, \&
  {Newman}}]{lin04}
{Lin}, L., {Koo}, D.~C., {Willmer}, C.~N.~A., {et~al.} 2004, \apjl, 617, L9

\bibitem[{{Lin} {et~al.}(2008){Lin}, {Patton}, {Koo}, {Casteels}, {Conselice},
  {Faber}, {Lotz}, {Willmer}, {Hsieh}, {Chiueh}, {Newman}, {Novak}, {Weiner},
  \& {Cooper}}]{lin08}
{Lin}, L., {Patton}, D.~R., {Koo}, D.~C., {et~al.} 2008, \apj, 681, 232

\bibitem[{{Lintott} {et~al.}(2008){Lintott}, {Schawinski}, {Slosar}, {Land},
  {Bamford}, {Thomas}, {Raddick}, {Nichol}, {Szalay}, {Andreescu}, {Murray}, \&
  {Vandenberg}}]{lintott08}
{Lintott}, C.~J., {Schawinski}, K., {Slosar}, A., {et~al.} 2008, \mnras, 389,
  1179

\bibitem[{{Liske} {et~al.}(2003){Liske}, {Lemon}, {Driver}, {Cross}, \&
  {Couch}}]{liske03}
{Liske}, J., {Lemon}, D.~J., {Driver}, S.~P., {Cross}, N.~J.~G., \& {Couch},
  W.~J. 2003, \mnras, 344, 307

\bibitem[{{L{\'o}pez-Sanjuan} {et~al.}(2009{\natexlab{a}}){L{\'o}pez-Sanjuan},
  {Balcells}, {Garc{\'{\i}}a-Dab{\'o}}, {Prieto}, {Crist{\'o}bal-Hornillos},
  {Eliche-Moral}, {Abreu}, {Erwin}, \& {Guzm{\'a}n}}]{clsj09ffgs}
{L{\'o}pez-Sanjuan}, C., {Balcells}, M., {Garc{\'{\i}}a-Dab{\'o}}, C.~E.,
  {et~al.} 2009{\natexlab{a}}, \apj, 694, 643

\bibitem[{{L{\'o}pez-Sanjuan} {et~al.}(2010{\natexlab{a}}){L{\'o}pez-Sanjuan},
  {Balcells}, {P{\'e}rez-Gonz{\'a}lez}, {Barro}, {Gallego}, \&
  {Zamorano}}]{clsj10pargoods}
{L{\'o}pez-Sanjuan}, C., {Balcells}, M., {P{\'e}rez-Gonz{\'a}lez}, P.~G.,
  {et~al.} 2010{\natexlab{a}}, \aap, 518, A20+

\bibitem[{{L{\'o}pez-Sanjuan} {et~al.}(2009{\natexlab{b}}){L{\'o}pez-Sanjuan},
  {Balcells}, {P{\'e}rez-Gonz{\'a}lez}, {Barro}, {Garc{\'{\i}}a-Dab{\'o}},
  {Gallego}, \& {Zamorano}}]{clsj09ffgoods}
{L{\'o}pez-Sanjuan}, C., {Balcells}, M., {P{\'e}rez-Gonz{\'a}lez}, P.~G.,
  {et~al.} 2009{\natexlab{b}}, \aap, 501, 505

\bibitem[{{L{\'o}pez-Sanjuan} {et~al.}(2010{\natexlab{b}}){L{\'o}pez-Sanjuan},
  {Balcells}, {P{\'e}rez-Gonz{\'a}lez}, {Barro}, {Garc{\'{\i}}a-Dab{\'o}},
  {Gallego}, \& {Zamorano}}]{clsj10megoods}
{L{\'o}pez-Sanjuan}, C., {Balcells}, M., {P{\'e}rez-Gonz{\'a}lez}, P.~G.,
  {et~al.} 2010{\natexlab{b}}, \apj, 710, 1170

\bibitem[{{Lotz} {et~al.}(2008){Lotz}, {Davis}, {Faber}, {Guhathakurta},
  {Gwyn}, {Huang}, {Koo}, {Le Floc'h}, {Lin}, {Newman}, {Noeske}, {Papovich},
  {Willmer}, {Coil}, {Conselice}, {Cooper}, {Hopkins}, {Metevier}, {Primack},
  {Rieke}, \& {Weiner}}]{lotz08ff}
{Lotz}, J.~M., {Davis}, M., {Faber}, S.~M., {et~al.} 2008, \apj, 672, 177

\bibitem[{{Lotz} {et~al.}(2010{\natexlab{a}}){Lotz}, {Jonsson}, {Cox}, \&
  {Primack}}]{lotz10gas}
{Lotz}, J.~M., {Jonsson}, P., {Cox}, T.~J., \& {Primack}, J.~R.
  2010{\natexlab{a}}, \mnras, 404, 590

\bibitem[{{Lotz} {et~al.}(2010{\natexlab{b}}){Lotz}, {Jonsson}, {Cox}, \&
  {Primack}}]{lotz10t}
{Lotz}, J.~M., {Jonsson}, P., {Cox}, T.~J., \& {Primack}, J.~R.
  2010{\natexlab{b}}, \mnras, 404, 575

\bibitem[{{Maller} {et~al.}(2006){Maller}, {Katz}, {Kere{\v s}}, {Dav{\'e}}, \&
  {Weinberg}}]{maller06}
{Maller}, A.~H., {Katz}, N., {Kere{\v s}}, D., {Dav{\'e}}, R., \& {Weinberg},
  D.~H. 2006, \apj, 647, 763

\bibitem[{{Martig} {et~al.}(2009){Martig}, {Bournaud}, {Teyssier}, \&
  {Dekel}}]{martig09}
{Martig}, M., {Bournaud}, F., {Teyssier}, R., \& {Dekel}, A. 2009, \apj, 707,
  250

\bibitem[{{Masters} {et~al.}(2011){Masters}, {Nichol}, {Hoyle}, {Lintott},
  {Bamford}, {Edmondson}, {Fortson}, {Keel}, {Schawinski}, {Smith}, \&
  {Thomas}}]{masters10}
{Masters}, K.~L., {Nichol}, R.~C., {Hoyle}, B., {et~al.} 2011, \mnras, in press
  [ArXiv: 1003.0449]

\bibitem[{{McCracken} {et~al.}(2003){McCracken}, {Radovich}, {Bertin},
  {Mellier}, {Dantel-Fort}, {Le F{\`e}vre}, {Cuillandre}, {Gwyn}, {Foucaud}, \&
  {Zamorani}}]{mccracken03}
{McCracken}, H.~J., {Radovich}, M., {Bertin}, E., {et~al.} 2003, \aap, 410, 17

\bibitem[{{Mihos} \& {Hernquist}(1994)}]{mihos94}
{Mihos}, J.~C. \& {Hernquist}, L. 1994, \apjl, 425, L13

\bibitem[{{Naab} {et~al.}(2006){Naab}, {Jesseit}, \& {Burkert}}]{naab06ss}
{Naab}, T., {Jesseit}, R., \& {Burkert}, A. 2006, \mnras, 372, 839

\bibitem[{{Naab} {et~al.}(2009){Naab}, {Johansson}, \& {Ostriker}}]{naab09}
{Naab}, T., {Johansson}, P.~H., \& {Ostriker}, J.~P. 2009, \apjl, 699, L178

\bibitem[{{Oesch} {et~al.}(2010){Oesch}, {Carollo}, {Feldmann}, {Hahn},
  {Lilly}, {Sargent}, {Scarlata}, {Aller}, {Aussel}, {Bolzonella}, {Bschorr},
  {Bundy}, {Capak}, {Ilbert}, {Kneib}, {Koekemoer}, {Kova{\v c}}, {Leauthaud},
  {Le Floc'h}, {Massey}, {McCracken}, {Pozzetti}, {Renzini}, {Rhodes},
  {Salvato}, {Sanders}, {Scoville}, {Sheth}, {Taniguchi}, \&
  {Thompson}}]{oesch10}
{Oesch}, P.~A., {Carollo}, C.~M., {Feldmann}, R., {et~al.} 2010, \apjl, 714,
  L47

\bibitem[{{Patton} \& {Atfield}(2008)}]{patton08}
{Patton}, D.~R. \& {Atfield}, J.~E. 2008, \apj, 685, 235

\bibitem[{{Patton} {et~al.}(2000){Patton}, {Carlberg}, {Marzke}, {Pritchet},
  {da Costa}, \& {Pellegrini}}]{patton00}
{Patton}, D.~R., {Carlberg}, R.~G., {Marzke}, R.~O., {et~al.} 2000, \apj, 536,
  153

\bibitem[{{Patton} {et~al.}(2011){Patton}, {Ellison}, {Simard}, {McConnachie},
  \& {Mendel}}]{patton11}
{Patton}, D.~R., {Ellison}, S.~L., {Simard}, L., {McConnachie}, A.~W., \&
  {Mendel}, J.~T. 2011, \mnras, 14

\bibitem[{{Patton} {et~al.}(2002){Patton}, {Pritchet}, {Carlberg}, {Marzke},
  {Yee}, {Hall}, {Lin}, {Morris}, {Sawicki}, {Shepherd}, \& {Wirth}}]{patton02}
{Patton}, D.~R., {Pritchet}, C.~J., {Carlberg}, R.~G., {et~al.} 2002, \apj,
  565, 208

\bibitem[{{Peng} {et~al.}(2010){Peng}, {Lilly}, {Kova{\v c}}, {Bolzonella},
  {Pozzetti}, {Renzini}, {Zamorani}, {Ilbert}, {Knobel}, {Iovino}, {Maier},
  {Cucciati}, {Tasca}, {Carollo}, {Silverman}, {Kampczyk}, {de Ravel},
  {Sanders}, {Scoville}, {Contini}, {Mainieri}, {Scodeggio}, {Kneib}, {Le
  F{\`e}vre}, {Bardelli}, {Bongiorno}, {Caputi}, {Coppa}, {de la Torre},
  {Franzetti}, {Garilli}, {Lamareille}, {Le Borgne}, {Le Brun}, {Mignoli},
  {Perez Montero}, {Pello}, {Ricciardelli}, {Tanaka}, {Tresse}, {Vergani},
  {Welikala}, {Zucca}, {Oesch}, {Abbas}, {Barnes}, {Bordoloi}, {Bottini},
  {Cappi}, {Cassata}, {Cimatti}, {Fumana}, {Hasinger}, {Koekemoer},
  {Leauthaud}, {Maccagni}, {Marinoni}, {McCracken}, {Memeo}, {Meneux}, {Nair},
  {Porciani}, {Presotto}, \& {Scaramella}}]{peng10}
{Peng}, Y., {Lilly}, S.~J., {Kova{\v c}}, K., {et~al.} 2010, \apj, 721, 193

\bibitem[{{Pozzetti} {et~al.}(2007){Pozzetti}, {Bolzonella}, {Lamareille},
  {Zamorani}, {Franzetti}, {Le F{\`e}vre}, {Iovino}, {Temporin}, {Ilbert},
  {Arnouts}, {Charlot}, {Brinchmann}, {Zucca}, {Tresse}, {Scodeggio}, {Guzzo},
  {Bottini}, {Garilli}, {Le Brun}, {Maccagni}, {Picat}, {Scaramella},
  {Vettolani}, {Zanichelli}, {Adami}, {Bardelli}, {Cappi}, {Ciliegi},
  {Contini}, {Foucaud}, {Gavignaud}, {McCracken}, {Marano}, {Marinoni},
  {Mazure}, {Meneux}, {Merighi}, {Paltani}, {Pell{\`o}}, {Pollo}, {Radovich},
  {Bondi}, {Bongiorno}, {Cucciati}, {de la Torre}, {Gregorini}, {Mellier},
  {Merluzzi}, {Vergani}, \& {Walcher}}]{pozzetti07}
{Pozzetti}, L., {Bolzonella}, M., {Lamareille}, F., {et~al.} 2007, \aap, 474,
  443

\bibitem[{{Pozzetti} {et~al.}(2010){Pozzetti}, {Bolzonella}, {Zucca},
  {Zamorani}, {Lilly}, {Renzini}, {Moresco}, {Mignoli}, {Cassata}, {Tasca},
  {Lamareille}, {Maier}, {Meneux}, {Halliday}, {Oesch}, {Vergani}, {Caputi},
  {Kova{\v c}}, {Cimatti}, {Cucciati}, {Iovino}, {Peng}, {Carollo}, {Contini},
  {Kneib}, {Le F{\'e}vre}, {Mainieri}, {Scodeggio}, {Bardelli}, {Bongiorno},
  {Coppa}, {de la Torre}, {de Ravel}, {Franzetti}, {Garilli}, {Kampczyk},
  {Knobel}, {Le Borgne}, {Le Brun}, {Pell{\`o}}, {Perez Montero},
  {Ricciardelli}, {Silverman}, {Tanaka}, {Tresse}, {Abbas}, {Bottini}, {Cappi},
  {Guzzo}, {Koekemoer}, {Leauthaud}, {Maccagni}, {Marinoni}, {McCracken},
  {Memeo}, {Porciani}, {Scaramella}, {Scarlata}, \& {Scoville}}]{pozzetti10}
{Pozzetti}, L., {Bolzonella}, M., {Zucca}, E., {et~al.} 2010, \aap, 523, A13+

\bibitem[{{Rawat} {et~al.}(2008){Rawat}, {Hammer}, {Kembhavi}, \&
  {Flores}}]{rawat08}
{Rawat}, A., {Hammer}, F., {Kembhavi}, A.~K., \& {Flores}, H. 2008, \apj, 681,
  1089

\bibitem[{{Robaina} {et~al.}(2010){Robaina}, {Bell}, {van der Wel},
  {Somerville}, {Skelton}, {McIntosh}, {Meisenheimer}, \& {Wolf}}]{robaina10}
{Robaina}, A.~R., {Bell}, E.~F., {van der Wel}, A., {et~al.} 2010, \apj, 719,
  844

\bibitem[{{Roseboom} {et~al.}(2006){Roseboom}, {Pimbblet}, {Drinkwater},
  {Cannon}, {de Propris}, {Edge}, {Eisenstein}, {Nichol}, {Smail}, {Wake},
  {Bland-Hawthorn}, {Bridges}, {Carson}, {Colless}, {Couch}, {Croom}, {Driver},
  {Hewett}, {Loveday}, {Ross}, {Schneider}, {Shanks}, {Sharp}, \&
  {Weilbacher}}]{roseboom06}
{Roseboom}, I.~G., {Pimbblet}, K.~A., {Drinkwater}, M.~J., {et~al.} 2006,
  \mnras, 373, 349

\bibitem[{{Rothberg} \& {Fischer}(2010)}]{rothberg10}
{Rothberg}, B. \& {Fischer}, J. 2010, \apj, 712, 318

\bibitem[{{Rothberg} \& {Joseph}(2006{\natexlab{a}})}]{rothberg06a}
{Rothberg}, B. \& {Joseph}, R.~D. 2006{\natexlab{a}}, \aj, 131, 185

\bibitem[{{Rothberg} \& {Joseph}(2006{\natexlab{b}})}]{rothberg06b}
{Rothberg}, B. \& {Joseph}, R.~D. 2006{\natexlab{b}}, \aj, 132, 976

\bibitem[{{Salim} {et~al.}(2009){Salim}, {Dickinson}, {Michael Rich},
  {Charlot}, {Lee}, {Schiminovich}, {P{\'e}rez-Gonz{\'a}lez}, {Ashby},
  {Papovich}, {Faber}, {Ivison}, {Frayer}, {Walton}, {Weiner}, {Chary},
  {Bundy}, {Noeske}, \& {Koekemoer}}]{salim09}
{Salim}, S., {Dickinson}, M., {Michael Rich}, R., {et~al.} 2009, \apj, 700, 161

\bibitem[{{Scarlata} {et~al.}(2007){Scarlata}, {Carollo}, {Lilly}, {Feldmann},
  {Kampczyk}, {Renzini}, {Cimatti}, {Halliday}, {Daddi}, {Sargent},
  {Koekemoer}, {Scoville}, {Kneib}, {Leauthaud}, {Massey}, {Rhodes}, {Tasca},
  {Capak}, {McCracken}, {Mobasher}, {Taniguchi}, {Thompson}, {Ajiki}, {Aussel},
  {Murayama}, {Sanders}, {Sasaki}, {Shioya}, \& {Takahashi}}]{scarlata07ee}
{Scarlata}, C., {Carollo}, C.~M., {Lilly}, S.~J., {et~al.} 2007, \apjs, 172,
  494

\bibitem[{{Schiminovich} {et~al.}(2007){Schiminovich}, {Wyder}, {Martin},
  {Johnson}, {Salim}, {Seibert}, {Treyer}, {Budav{\'a}ri}, {Hoopes},
  {Zamojski}, {Barlow}, {Forster}, {Friedman}, {Morrissey}, {Neff}, {Small},
  {Bianchi}, {Donas}, {Heckman}, {Lee}, {Madore}, {Milliard}, {Rich}, {Szalay},
  {Welsh}, \& {Yi}}]{schi07}
{Schiminovich}, D., {Wyder}, T.~K., {Martin}, D.~C., {et~al.} 2007, \apjs, 173,
  315

\bibitem[{{Scodeggio} {et~al.}(2005){Scodeggio}, {Franzetti}, {Garilli},
  {Zanichelli}, {Paltani}, {Maccagni}, {Bottini}, {Le Brun}, {Contini},
  {Scaramella}, {Adami}, {Bardelli}, {Zucca}, {Tresse}, {Ilbert}, {Foucaud},
  {Iovino}, {Merighi}, {Zamorani}, {Gavignaud}, {Rizzo}, {McCracken}, {Le
  F{\`e}vre}, {Picat}, {Vettolani}, {Arnaboldi}, {Arnouts}, {Bolzonella},
  {Cappi}, {Charlot}, {Ciliegi}, {Guzzo}, {Marano}, {Marinoni}, {Mathez},
  {Mazure}, {Meneux}, {Pell{\`o}}, {Pollo}, {Pozzetti}, \&
  {Radovich}}]{scodeggio05}
{Scodeggio}, M., {Franzetti}, P., {Garilli}, B., {et~al.} 2005, \pasp, 117,
  1284

\bibitem[{{Scoville} {et~al.}(2007){Scoville}, {Aussel}, {Brusa}, {Capak},
  {Carollo}, {Elvis}, {Giavalisco}, {Guzzo}, {Hasinger}, {Impey}, {Kneib},
  {LeFevre}, {Lilly}, {Mobasher}, {Renzini}, {Rich}, {Sanders}, {Schinnerer},
  {Schminovich}, {Shopbell}, {Taniguchi}, \& {Tyson}}]{scoville07}
{Scoville}, N., {Aussel}, H., {Brusa}, M., {et~al.} 2007, \apjs, 172, 1

\bibitem[{{Sheth} {et~al.}(2008){Sheth}, {Elmegreen}, {Elmegreen}, {Capak},
  {Abraham}, {Athanassoula}, {Ellis}, {Mobasher}, {Salvato}, {Schinnerer},
  {Scoville}, {Spalsbury}, {Strubbe}, {Carollo}, {Rich}, \& {West}}]{sheth08}
{Sheth}, K., {Elmegreen}, D.~M., {Elmegreen}, B.~G., {et~al.} 2008, \apj, 675,
  1141

\bibitem[{{Springel} {et~al.}(2005){Springel}, {White}, {Jenkins}, {Frenk},
  {Yoshida}, {Gao}, {Navarro}, {Thacker}, {Croton}, {Helly}, {Peacock}, {Cole},
  {Thomas}, {Couchman}, {Evrard}, {Colberg}, \& {Pearce}}]{springel05}
{Springel}, V., {White}, S.~D.~M., {Jenkins}, A., {et~al.} 2005, \nat, 435, 629

\bibitem[{{Stewart} {et~al.}(2009){Stewart}, {Bullock}, {Wechsler}, \&
  {Maller}}]{stewart09}
{Stewart}, K.~R., {Bullock}, J.~S., {Wechsler}, R.~H., \& {Maller}, A.~H. 2009,
  \apj, 702, 307

\bibitem[{{Taylor} {et~al.}(2010){Taylor}, {Franx}, {Glazebrook}, {Brinchmann},
  {van der Wel}, \& {van Dokkum}}]{taylor10}
{Taylor}, E.~N., {Franx}, M., {Glazebrook}, K., {et~al.} 2010, \apj, 720, 723

\bibitem[{{Toft} {et~al.}(2009){Toft}, {Franx}, {van Dokkum}, {F{\"o}rster
  Schreiber}, {Labbe}, {Wuyts}, \& {Marchesini}}]{toft09}
{Toft}, S., {Franx}, M., {van Dokkum}, P., {et~al.} 2009, \apj, 705, 255

\bibitem[{{Trujillo} {et~al.}(2009){Trujillo}, {Cenarro}, {de
  Lorenzo-C{\'a}ceres}, {Vazdekis}, {de la Rosa}, \& {Cava}}]{trujillo09}
{Trujillo}, I., {Cenarro}, A.~J., {de Lorenzo-C{\'a}ceres}, A., {et~al.} 2009,
  \apjl, 692, L118

\bibitem[{{Trujillo} {et~al.}(2007){Trujillo}, {Conselice}, {Bundy}, {Cooper},
  {Eisenhardt}, \& {Ellis}}]{trujillo07}
{Trujillo}, I., {Conselice}, C.~J., {Bundy}, K., {et~al.} 2007, \mnras, 382,
  109

\bibitem[{{Trujillo} {et~al.}(2006){Trujillo}, {F{\"o}rster Schreiber},
  {Rudnick}, {Barden}, {Franx}, {Rix}, {Caldwell}, {McIntosh}, {Toft},
  {H{\"a}ussler}, {Zirm}, {van Dokkum}, {Labb{\'e}}, {Moorwood},
  {R{\"o}ttgering}, {van der Wel}, {van der Werf}, \& {van
  Starkenburg}}]{trujillo06}
{Trujillo}, I., {F{\"o}rster Schreiber}, N.~M., {Rudnick}, G., {et~al.} 2006,
  \apj, 650, 18

\bibitem[{{van der Wel}(2008)}]{vanderwel08}
{van der Wel}, A. 2008, \apjl, 675, L13

\bibitem[{{van der Wel} {et~al.}(2008){van der Wel}, {Holden}, {Zirm}, {Franx},
  {Rettura}, {Illingworth}, \& {Ford}}]{vanderwel08esize}
{van der Wel}, A., {Holden}, B.~P., {Zirm}, A.~W., {et~al.} 2008, \apj, 688, 48

\bibitem[{{van Dokkum} {et~al.}(2008){van Dokkum}, {Franx}, {Kriek}, {Holden},
  {Illingworth}, {Magee}, {Bouwens}, {Marchesini}, {Quadri}, {Rudnick},
  {Taylor}, \& {Toft}}]{vandokkum08}
{van Dokkum}, P.~G., {Franx}, M., {Kriek}, M., {et~al.} 2008, \apjl, 677, L5

\bibitem[{{van Dokkum} {et~al.}(2010){van Dokkum}, {Whitaker}, {Brammer},
  {Franx}, {Kriek}, {Labb{\'e}}, {Marchesini}, {Quadri}, {Bezanson},
  {Illingworth}, {Muzzin}, {Rudnick}, {Tal}, \& {Wake}}]{vandokkum10}
{van Dokkum}, P.~G., {Whitaker}, K.~E., {Brammer}, G., {et~al.} 2010, \apj,
  709, 1018

\bibitem[{{Vergani} {et~al.}(2008){Vergani}, {Scodeggio}, {Pozzetti}, {Iovino},
  {Franzetti}, {Garilli}, {Zamorani}, {Maccagni}, {Lamareille}, {Le F{\`e}vre},
  {Charlot}, {Contini}, {Guzzo}, {Bottini}, {Le Brun}, {Picat}, {Scaramella},
  {Tresse}, {Vettolani}, {Zanichelli}, {Adami}, {Arnouts}, {Bardelli},
  {Bolzonella}, {Cappi}, {Ciliegi}, {Foucaud}, {Gavignaud}, {Ilbert},
  {McCracken}, {Marano}, {Marinoni}, {Mazure}, {Meneux}, {Merighi}, {Paltani},
  {Pell{\`o}}, {Pollo}, {Radovich}, {Zucca}, {Bondi}, {Bongiorno},
  {Brinchmann}, {Cucciati}, {de la Torre}, {Gregorini}, {Perez-Montero},
  {Mellier}, {Merluzzi}, \& {Temporin}}]{vergani08}
{Vergani}, D., {Scodeggio}, M., {Pozzetti}, L., {et~al.} 2008, \aap, 487, 89

\bibitem[{{Walcher} {et~al.}(2008){Walcher}, {Lamareille}, {Vergani},
  {Arnouts}, {Buat}, {Charlot}, {Tresse}, {Le F{\`e}vre}, {Bolzonella},
  {Brinchmann}, {Pozzetti}, {Zamorani}, {Bottini}, {Garilli}, {Le Brun},
  {Maccagni}, {Milliard}, {Scaramella}, {Scodeggio}, {Vettolani}, {Zanichelli},
  {Adami}, {Bardelli}, {Cappi}, {Ciliegi}, {Contini}, {Franzetti}, {Foucaud},
  {Gavignaud}, {Guzzo}, {Ilbert}, {Iovino}, {McCracken}, {Marano}, {Marinoni},
  {Mazure}, {Meneux}, {Merighi}, {Paltani}, {Pell{\`o}}, {Pollo}, {Radovich},
  {Zucca}, {Lonsdale}, \& {Martin}}]{walcher08}
{Walcher}, C.~J., {Lamareille}, F., {Vergani}, D., {et~al.} 2008, \aap, 491,
  713

\bibitem[{{Wild} {et~al.}(2009){Wild}, {Walcher}, {Johansson}, {Tresse},
  {Charlot}, {Pollo}, {Le F{\`e}vre}, \& {de Ravel}}]{wild09}
{Wild}, V., {Walcher}, C.~J., {Johansson}, P.~H., {et~al.} 2009, \mnras, 395,
  144

\bibitem[{{Williams} {et~al.}(2009){Williams}, {Quadri}, {Franx}, {van Dokkum},
  \& {Labb{\'e}}}]{williams09}
{Williams}, R.~J., {Quadri}, R.~F., {Franx}, M., {van Dokkum}, P., \&
  {Labb{\'e}}, I. 2009, \apj, 691, 1879

\bibitem[{{Williams} {et~al.}(2010){Williams}, {Quadri}, {Franx}, {van Dokkum},
  {Toft}, {Kriek}, \& {Labb{\'e}}}]{williams10}
{Williams}, R.~J., {Quadri}, R.~F., {Franx}, M., {et~al.} 2010, \apj, 713, 738

\bibitem[{{Woods} \& {Geller}(2007)}]{woods07}
{Woods}, D.~F. \& {Geller}, M.~J. 2007, \aj, 134, 527

\bibitem[{{Wyder} {et~al.}(2007){Wyder}, {Martin}, {Schiminovich}, {Seibert},
  {Budav{\'a}ri}, {Treyer}, {Barlow}, {Forster}, {Friedman}, {Morrissey},
  {Neff}, {Small}, {Bianchi}, {Donas}, {Heckman}, {Lee}, {Madore}, {Milliard},
  {Rich}, {Szalay}, {Welsh}, \& {Yi}}]{wyder07}
{Wyder}, T.~K., {Martin}, D.~C., {Schiminovich}, D., {et~al.} 2007, \apjs, 173,
  293

\bibitem[{{Zheng} {et~al.}(2007){Zheng}, {Bell}, {Papovich}, {Wolf},
  {Meisenheimer}, {Rix}, {Rieke}, \& {Somerville}}]{zheng07}
{Zheng}, X.~Z., {Bell}, E.~F., {Papovich}, C., {et~al.} 2007, \apjl, 661, L41

\bibitem[{{Zucca} {et~al.}(2009){Zucca}, {Bardelli}, {Bolzonella}, {Zamorani},
  {Ilbert}, {Pozzetti}, {Mignoli}, {Kova{\v c}}, {Lilly}, {Tresse}, {Tasca},
  {Cassata}, {Halliday}, {Vergani}, {Caputi}, {Carollo}, {Contini}, {Kneib},
  {Le F{\`e}vre}, {Mainieri}, {Renzini}, {Scodeggio}, {Bongiorno}, {Coppa},
  {Cucciati}, {de La Torre}, {de Ravel}, {Franzetti}, {Garilli}, {Iovino},
  {Kampczyk}, {Knobel}, {Lamareille}, {Le Borgne}, {Le Brun}, {Maier},
  {Pell{\`o}}, {Peng}, {Perez-Montero}, {Ricciardelli}, {Silverman}, {Tanaka},
  {Abbas}, {Bottini}, {Cappi}, {Cimatti}, {Guzzo}, {Koekemoer}, {Leauthaud},
  {Maccagni}, {Marinoni}, {McCracken}, {Memeo}, {Meneux}, {Moresco}, {Oesch},
  {Porciani}, {Scaramella}, {Arnouts}, {Aussel}, {Capak}, {Kartaltepe},
  {Salvato}, {Sanders}, {Scoville}, {Taniguchi}, \& {Thompson}}]{zucca09}
{Zucca}, E., {Bardelli}, S., {Bolzonella}, M., {et~al.} 2009, \aap, 508, 1217

\bibitem[{{Zucca} {et~al.}(2006){Zucca}, {Ilbert}, {Bardelli}, {Tresse},
  {Zamorani}, {Arnouts}, {Pozzetti}, {Bolzonella}, {McCracken}, {Bottini},
  {Garilli}, {Le Brun}, {Le F{\`e}vre}, {Maccagni}, {Picat}, {Scaramella},
  {Scodeggio}, {Vettolani}, {Zanichelli}, {Adami}, {Arnaboldi}, {Cappi},
  {Charlot}, {Ciliegi}, {Contini}, {Foucaud}, {Franzetti}, {Gavignaud},
  {Guzzo}, {Iovino}, {Marano}, {Marinoni}, {Mazure}, {Meneux}, {Merighi},
  {Paltani}, {Pell{\`o}}, {Pollo}, {Radovich}, {Bondi}, {Bongiorno},
  {Busarello}, {Cucciati}, {Gregorini}, {Lamareille}, {Mathez}, {Mellier},
  {Merluzzi}, {Ripepi}, \& {Rizzo}}]{zucca06}
{Zucca}, E., {Ilbert}, O., {Bardelli}, S., {et~al.} 2006, \aap, 455, 879

\end{thebibliography}
\bibliographystyle{aa}

\begin{appendix}
\section{Merger fraction fitting by Generalized Least Squares}\label{mcfit}
The dependence of the merger fraction $f_{\rm m}\,(\geq \mu)$ with $\mu$ is well described by a power-law function (Eq.~[\ref{ffmueq}]). However, our definition of $f_{\rm m}$ is cumulative, so the points in Tables~\ref{ffmutab}, \ref{ffmuredtab} and \ref{ffmubluetab} are not independent and their errors are correlated. To obtain reliable fit parameters and their uncertainties we used the Generalized Least Squares (GLS; \citealt{aitken34}) method, which takes into account not only the variance of the data, but also the covariance between them. For a given $r_{\rm p}^{\rm max}$ and redshift range, we followed the next steps to estimate the covariance matrix of the data:
\begin{enumerate}
\item We extracted a random point, named $f_{\rm m}^{\rm sim}\,(\mu \geq 1/2)$, as drawn for a Gaussian distribution with mean $f_{\rm m}\,(\mu \geq 1/2)$ and standard deviation $\sigma_{f_{\rm m}}(\mu \geq 1/2)$. In this process we imposed that the random point had to be positive, i.e., negative merger fractions are nonphysical.

\item To obtain the next merger fraction, named $f_{\rm m}^{\rm sim}\,(\mu \geq 1/3)$, we extracted a random point as drawn for a Gaussian with mean $f_{\rm m}\,(\mu \geq 1/3) - f_{\rm m}\,(\mu \geq 1/2)$ and standard deviation $[\sigma^{2}_{f_{\rm m}}(\mu \geq 1/3) - \sigma^{2}_{f_{\rm m}}(\mu \geq 1/2)]^{1/2}$, and added it to the previous $f_{\rm m}^{\rm sim}\,(\mu > 1/2)$. In this process we set a negative random point to zero, that is, we imposed that merger fractions are cumulative when $\mu$ decreases. In addition, this process takes into account that the errors are correlated.

\item We repeated the step~2 for all the $\mu$ values under study down to $\mu = 1/10$. This provided us a set of $f_{\rm m}^{\rm sim}\,(\geq \mu)$.

\item We repeated 100000 times the steps~$1-3$ and estimated the covariance matrix of the observational merger fractions using the simulated ones.
\end{enumerate}

\begin{figure}[t!]
\resizebox{\hsize}{!}{\includegraphics{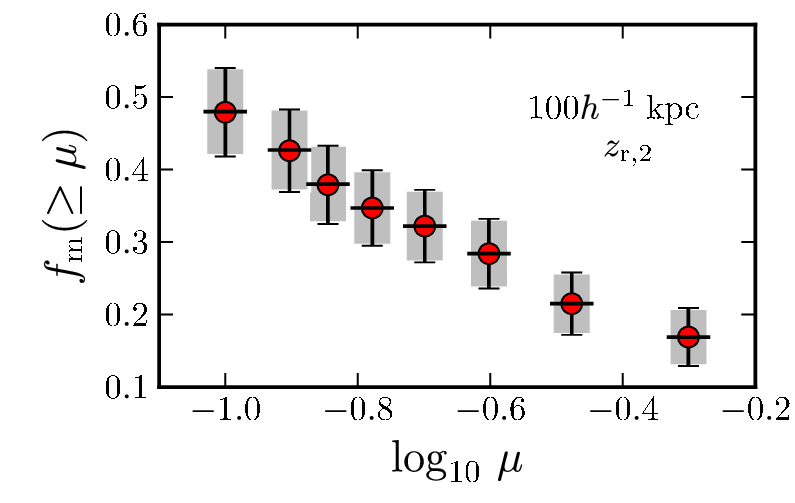}}
\resizebox{\hsize}{!}{\includegraphics{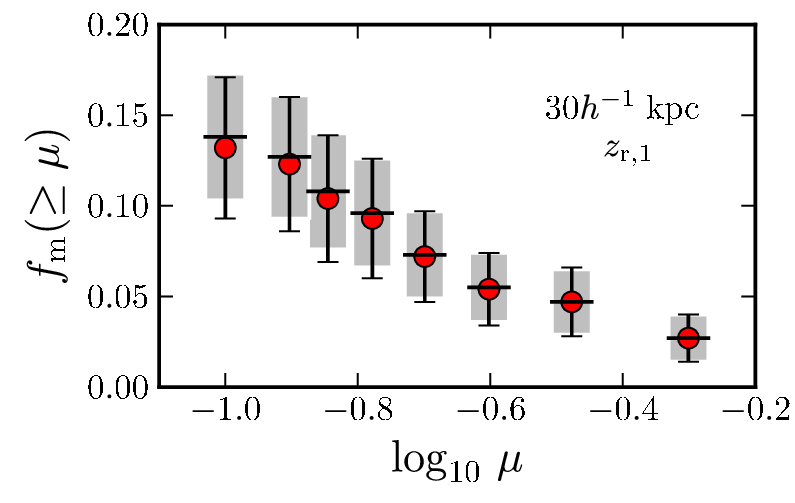}}
\caption{Merger fraction as a function of $\log_{10}\,\mu$. We use these particular axis to facilitate the visualization. The dots and the error bars are the observational data. The gray areas are the 1$\sigma$ confidence intervals of the simulated merger fractions, while the horizontal black lines are their mean (see text for details). {\it Top panel}: Merger fraction for $r_{\rm p}^{\rm max} = 100h^{-1}$ kpc at $z_{\rm r,2}$. {\it Bottom panel}: Merger fraction for $r_{\rm p}^{\rm max} = 30h^{-1}$ kpc at $z_{\rm r,1}$. [{\it A colour version of this plot is available at the electronic edition}].}
\label{ffsimfig}
\end{figure}

\begin{figure}[t!]
\resizebox{\hsize}{!}{\includegraphics{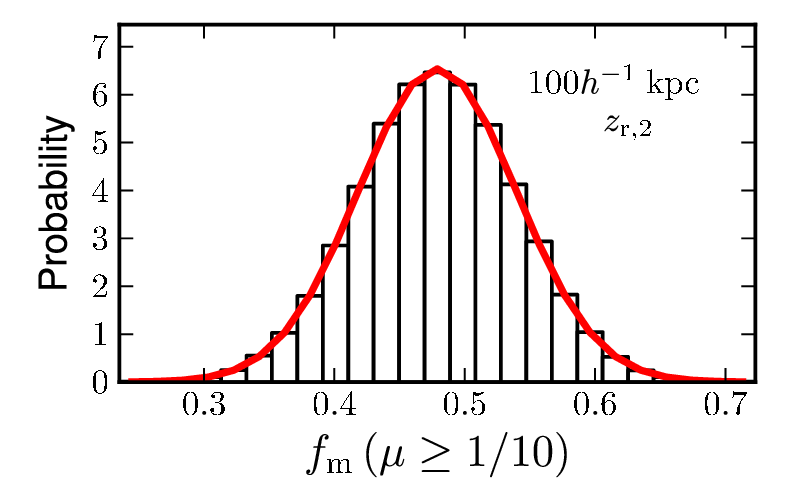}}
\resizebox{\hsize}{!}{\includegraphics{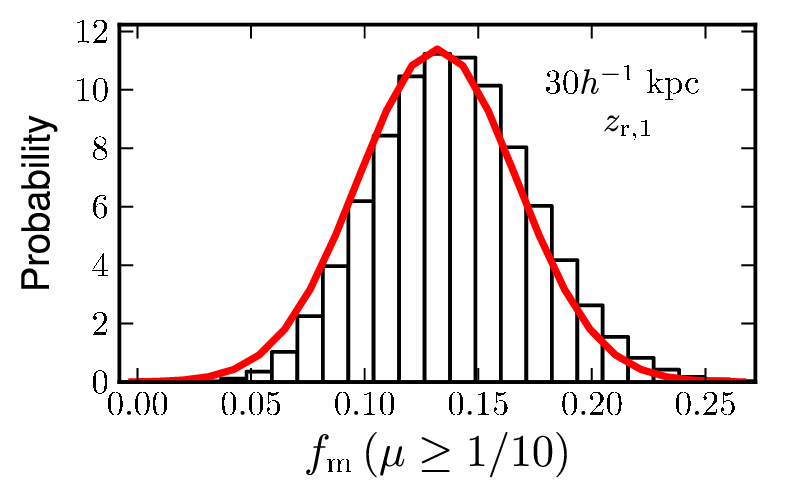}}
\caption{Probability distributions of the simulated merger fractions for $\mu > 1/10$ and $r_{\rm p}^{\rm max} = 100h^{-1}$ at $z_{\rm r,2}$ ({\it upper}), and for $r_{\rm p}^{\rm max} = 30h^{-1}$ at $z_{\rm r,1}$ ({\it lower}). The solid line is not a fit to the histogram, but the expected distribution from the observational merger fraction. [{\it A colour version of this plot is available at the electronic edition}].}
\label{gaussmcfit}
\end{figure}

We checked that our simulated merger fractions are a good description of the observational ones. We found that all the distributions of $f_{\rm m}^{\rm sim}\,(\geq \mu)$ are well described by a Gaussian, as desired. In Fig.~\ref{ffsimfig} we show the observational and the simulated merger fractions for $r_{\rm p}^{\rm max} = 100h^{-1}$ kpc at $z_{\rm r,2}$ and for $r_{\rm p}^{\rm max} = 30h^{-1}$ kpc at $z_{\rm r,1}$. We choose these two examples because they are the best and the worst simulated cases, respectively. Observational and simulated merger fractions are in agreement in the first case, but in the second case the values of the merger fraction are slightly overestimated (less than 5\%), while the standard deviations are underestimated (less than 10\%). To understand the origin of this discrepancy, we studied the distribution of $f_{\rm m}^{\rm sim}\,(\mu \geq 1/10)$ for both cases, Fig.~\ref{gaussmcfit}. In the first case the simulated distribution and that expected from observations are in excellent agreement. However, in the second case we find less points than expected at low values of the merger fraction. This is due to the lower values of the observed merger fraction at $z_{\rm r, 1}$ and the higher errors for $r_{\rm p}^{\rm max} = 30h^{-1}$ kpc measurements. This leads to negative random points, which we did not take into account (step~1) or set to zero (step~2), so we missed simulated values in the lower tail of the distribution. Despite of that, the global simulated distribution is a good description of the expected one: if we only use the upper tail of the distribution to describe it, the difference between the observed and the simulated values of the merger fraction and its standard deviation becomes lower than 2\% and 3\%, respectively. Hence, we conclude that the simulated merger fractions describe well the observational ones and that the estimated covariance matrix is a good approximation to the real one.

\begin{figure}[t!]
\resizebox{\hsize}{!}{\includegraphics{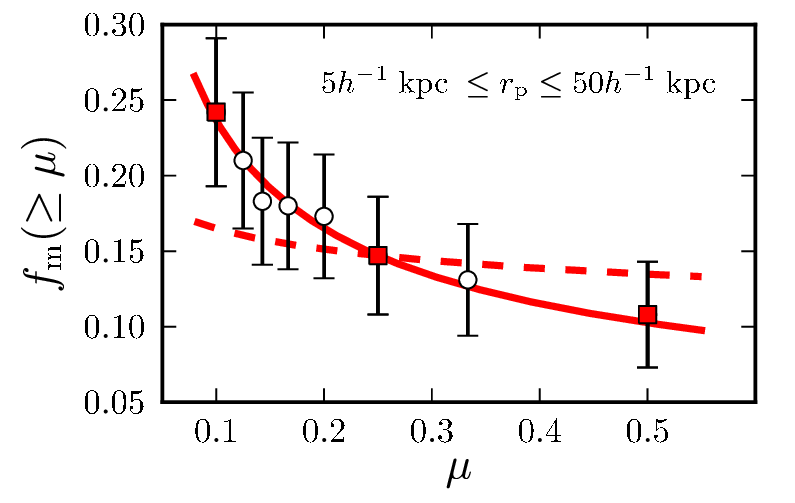}}
\caption{Generalized Least Squares fit to all the data (dashed line) and to the squares ($\mu = 1/10, 1/4$ and 1/2; solid line). Observed merger fractions are for $r_{\rm p}^{\rm max} = 50h^{-1}$ kpc at $z_{\rm r,2}$. [{\it A colour version of this plot is available at the electronic edition}].}
\label{2fitfig}
\end{figure}

Using the covariance matrix, we applied the GLS to estimate $f_{\rm MM}$ and $s$ (Eq.~[\ref{ffmueq}]). We noted that the errors in $f_{\rm MM}$ are similar or higher than the errors in the observed major merger fractions, so we can not obtain new information of $f_{\rm MM}$ from the GLS analysis. Hence, we set the value of $f_{\rm MM}$ to the observed one and used GLS to estimate the power-law index $s$. To obtain reliable fits given the cumulative nature of the data, we opted to use the $\mu = 1/10$ (lower $\mu$ value), $\mu = 1/4$ (the fixed major merger fraction) and $\mu = 1/2$ (higher $\mu$ value) data points, as we noticed that, as expected, most of the slope information is contained in these three points \citep{jaech64}. This produces a stable fit at every $r_{\rm p}^{\rm max}$, as shown in Fig.~\ref{2fitfig} for $50h^{-1}$ kpc separations. Adding other 5 intermediate points is only decreasing the variance on $s$ by 10-15\% but is producing low quality fits as shown in Fig.~\ref{2fitfig} (i.e., the fitted curves depart more than 1$\sigma$ from the observational data), which is traced to the increase in observational errors: analytically all the information is contained in a few $\mu$ points and the GLS does not take into account most of the data in the fit.

In summary, all the power-law index $s$ quoted in the paper were obtained from a GLS fit to $\mu = 1/10, 1/4$ and 1/2 merger fraction data, and using simulated merger fractions to estimate their covariance matrix.

\end{appendix}

\end{document}